\documentstyle[twocolumn,aps,epsfig,amsfonts]{revtex}

\newcommand{\pprime}{{\prime\prime}}
\newcommand{\bra}{\langle}
\newcommand{\ket}{\rangle}
\newcommand{\order}{{\cal O}}
\newcommand{\bOmega}{\bbox{\Omega}}
\newcommand{\bxi}{\bbox{\xi}}
\newcommand{\bomega}{\bbox{\omega}}
\newcommand{\bpsi}{\bbox{\psi}}
\newcommand{\hC}{\widehat{C}}

\newcommand{\hL}{\widehat{L}}
\newcommand{\hK}{\widehat{K}}
\newcommand{\bR}{\bbox{R}}
\newcommand{\sgn}{\textrm{sgn}}
\newcommand{\erf}{\textrm{erf}}
\newcommand{\Tr}{\textrm{Tr}}
\newcommand{\bq}{\bbox{q}}
\newcommand{\hq}{\widehat{q}}

\newcommand{\bw}{\bbox{w}}
\newcommand{\bx}{\bbox{x}}
\newcommand{\hw}{\widehat{w}}
\newcommand{\hx}{\widehat{x}}

\newcommand{\NN}{\mathbb{N}}

\newtheorem{lemma}{Lemma}
\newenvironment{proof}{\textbf{Proof~~}}{\hfill\rule{6pt}{6pt}}

\begin{document}
\draft

\title{Generating Functional Analysis of the Dynamics of the\\ Batch Minority Game with Random External Information}
\author{J.A.F. Heimel and A.C.C. Coolen}
\address{Department of Mathematics, King's College London, The Strand, London WC2R 2LS, UK}
\date{\today}

\maketitle

\begin{abstract}
We study the dynamics of the batch minority game, with random external
information,
using generating functional techniques \`{a} la De Dominicis. The relevant
control parameter in this model is the ratio $\alpha=p/N$ of the
number $p$ of possible values for the external information over
the number $N$ of trading agents. In the limit $N\to\infty$ we calculate the
location $\alpha_c$ of the phase transition (signaling  the onset of anomalous response),
and solve the statics for $\alpha>\alpha_c$ exactly.
The temporal correlations in global market
fluctuations turn out not to decay to zero for infinitely widely separated
times. For $\alpha<\alpha_c$ the stationary state is shown
to be non-unique. For $\alpha\to 0$
we analyse our equations in leading order in
$\alpha$, and find asymptotic solutions with diverging volatility
$\sigma=\order(\alpha^{-\frac{1}{2}})$ (as
regularly observed in simulations), but also asymptotic solutions with
vanishing volatility $\sigma=\order(\alpha^{\frac{1}{2}})$. The former, however, are shown to
emerge only if the agents' initial strategy valuations are below a specific
critical value.
\end{abstract}

\narrowtext

\section{Introduction}

The minority game has been the subject of much (and at times
heated) debate in the physics literature recently. It was originally introduced
in \cite{ChalZhan97}, as a variation of the El Farrol-Bar problem \cite{Arth94},
to serve as a simple model for a market in economics, and has since then attracted much
attention (see e.g. \cite{Chalweb}). The players in the minority game are trading agents
who, at every stage of the game, have to make a decision whether to buy or to sell,
on the basis of both publicly available information (i.e. past market
dynamics, weather forecasts, political developments or stock prices) and their personal strategies.
Those agents who find themselves having made the minority decision
 make a profit, while those agents which opted for the majority
choice loose money. After each round all agents re-value their
strategies. There are many variations on the precise implementation of this game,
yet most share the same main features of the emerging market fluctuations.
The important control parameter in the model is the ratio $\alpha=p/N$ of the
number $p$ of possible values for the external information over
the number $N$ of trading agents. If this ratio $\alpha$ is very large, the agents
exhibit essentially random behaviour. This is reflected in the fluctuations of the
total bid, being the sum of all buyers minus the sum of all sellers. If less
external information is available (or used) to base decisions upon, i.e. for reduced $\alpha$, the
mismatch between buyers and sellers is found to decrease, and the market behaves
more efficiently. This behaviour is now understood quite well on the basis of the replica
calculations in \cite{ChalMarsZecc00,MarsChalZecc00,DemaMars00} and the
crowd-anticrowd theory of \cite{JohnHartHui98}. The situation is much less clear, however,
when $\alpha$ becomes very small. One possibility is
that the market becomes extremely efficient, and the number of buyers
almost equals the number of sellers.
Another possibility is that the mismatch between buyers and sellers
diverges if the amount of shared (i.e. external) information becomes small, and the market becomes
extremely inefficient (see e.g. \cite{SaviManuRiol99,ManuLiRiolSavi98}).

In this paper we solve the dynamics
for the original many agent model,
using the exact generating functional (or path integral) techniques introduced in \cite{DeDo78}.
After defining the rules of the game we derive in the limit $N\to\infty$
an equivalent description in terms of an
effective stochastic non-Markovian single agent process, for which we calculate the first
time steps. For sufficiently large values of $\alpha$, we can  solve the statics
exactly under the assumption of absence of anomalous response. We calculate the point $\alpha_c$
where this assumption breaks down, resulting in a phase transition; our value for $\alpha_c$
is identical to that found in \cite{ChalMarsZecc00}.
The present dynamical approach allows us to study the behaviour of the market
below $\alpha_c$. In this region there exist persistent non-static
solutions which cannot be studied by the methods of \cite{ChalMarsZecc00}.
Below $\alpha_c$ the market is non-ergodic and 
the initial conditions of the agents determine
the final stationary state of the market.
For $\alpha\to 0$ we calculate the market volatility in leading order
in $\alpha$ for the case where the agents
are initialised with only weak strategy preferences, leading to a diverging volatility with exactly
the scaling exponent $\sigma=\order(\alpha^{-\frac{1}{2}})$ predicted in \cite{ManuLiRiolSavi98}
on the basis of
heuristic arguments. We find a critical value for the initial
strategy valuations above which this solution no longer
exists, and is being replaced by an alternative solution with a vanishing
volatility of the form $\sigma=\order(\alpha^{\frac{1}{2}})$.
Our dynamical approach allows in addition for the calculation of the two-time
correlations in the global market fluctuations,  by definition
inaccessible with equilibrium methods (replica or otherwise),
which are found
to have a persistent component.
Numerical simulations confirm our theoretical results convincingly.

\section{Model Definitions}

There are $N$ agents playing the game. We will only consider the case where $N$
is very large, and ultimately take the limit $N\to\infty$.
The agents are labeled with Roman indices $i,j,k,$ etc. At each iteration round
$\ell$ all agents are given the same (as yet unspecified)
piece of external information $I_{\mu(\ell)}$, chosen randomly from a total number
$p=\alpha N$ of possible values, i.e. $\mu(\ell)\in\{1,\ldots,\alpha N\}$.
In the original model \cite{ChalZhan97} the history of the actual market is used as the information
given to the agents.
Each agent $i$ has $S$ strategies $\bR_{ia}=(R_{ia}^1,\ldots,R_{ia}^{\alpha N})\in\{-1,1\}^{\alpha N}$
at her disposal with which to determine how to
convert the external information into a trading decision, with $a\in\{1,\ldots,S\}$.
Each component $R_{ia}^\mu$ is selected randomly and independently
from $\{-1,1\}$ before the start of the game, with uniform probabilities, and remains fixed throughout
the game. The strategies thus introduce quenched disorder into the model.
Each strategy of every agent is given an initial valuation or pay-off $p_{ia}(0)$.
The choice made for these initial values will turn out to be crucial for the emerging behaviour of the
market. Given a choice $\mu(\ell)$ made for the external information presented at
the start of round $\ell$, every agent $i$ selects the strategy $\tilde{a}_i(\ell)$
which for trader $i$ has the
highest pay-off value at that point in time, i.e. $\tilde{a}_i(\ell)=\mbox{arg max}~ p_{ia}(\ell)$, and
subsequently makes
a binary bid $b_i(\ell)=R^{\mu(\ell)}_{i\tilde{a}_i(t)}$. The
(re-scaled) total bid at stage $\ell$ is defined as
$A(\ell)=N^{-1/2}\sum_i b_i(\ell)$. Next all agents update the pay-off values of each
strategy $a$ on the basis of what would have happened if they had played
that particular strategy:
\[
  p_{ia}(\ell\!+\!1)=p_{ia}(\ell) - R^{\mu(\ell)}_{ia} A(\ell)
\]
The minus sign in this expression has the effect that strategies that would have produced a
minority decision are appreciated.

This setup so far allows for an arbitrary number of strategies $S$. The qualitative
behaviour of the market fluctuations,
however, is found to be very much the
same for all non-extensive number of strategies  larger than one
\cite{Cava99}. We therefore present results here only for the $S=2$ model,
where the equations can be simplified considerably upon
introducing for each agent the instantaneous difference between the two strategy valuations,
$q_i(\ell)=[p_{i1}(\ell)-p_{i2}(\ell)]/2$, as well as the
average strategy $\bomega_i=(\bR_{i1}+\bR_{i2})/2$ and
the difference between the
strategies
$\bxi_i=(\bR_{i1}-\bR_{i2})/2$. The actually selected strategy in round $\ell$ can now be written
explicitly as a function of $s_i(\ell)=\mbox{sgn}[q_i(\ell)]$,
viz. $\bR_{i\tilde{a}_i(\ell)}=\bomega_i+s_i(\ell)\bxi_i$, and the
evolution of the difference will now be given by:
\begin{equation}
\label{eq:online}
  q_i(\ell\!+\!1)=q_i(\ell)-\xi_i^{\mu(\ell)}[
    \Omega^{\mu(\ell)} \! + \!
    N^{-\frac{1}{2}}\sum_j \xi_j^{\mu(\ell)} s_j(\ell)
  ],
\end{equation}
with $\bOmega=N^{-1/2}\sum_j \bomega_j\in\Re^{\alpha N}$.
It has been observed in numerical simulations, see e.g.
\cite{GarrMoroSher00},
that the magnitude of the market fluctuations remains almost unchanged if a
large number of bids are performed before a re-evaluation of the strategies is carried out.
This is the motivation for us to study a modified (and simpler)
version of the dynamics of the game, where, rather than allowing the strategy pay-off valuations to
be changed at each round, only the accumulated effect of a
large number of market decisions is used to change an agent's strategy
pay-off valuations. This amounts to performing an average in the above dynamic equations
over the choices to be made for the external information. If we also change the time-unit
accordingly
from $\ell$ (which measured individual rounds of the game) to a
new unit $t$ which is proportional to the number of pay-off validation updates,
we arrive at
\begin{equation}\label{eq:batch}
  q_i(t+1)=q_i(t)- h_i - \sum_j J_{ij} s_j(t)
\end{equation}
where $J_{ij}=\bxi_i\cdot\bxi_j/N\tau^2$ and
$h_i=\bxi_i\cdot\bOmega/\sqrt{N}\tau^2$, and with
$\tau^2=\langle(\Omega^\mu)^2\rangle=\langle(\xi_i^\mu)^2\rangle=\langle(\omega_i^\mu)^2\rangle$;
here $\tau^2=\frac{1}{2})$.  The above particular choice of time
scaling has been made only in view of it giving the simplest equations later.
To make a connection with the original game, one must interpret the
evolution of the $q_i(t)$ as described by (\ref{eq:batch}) as the
accumulated effect of order $N$ iterations in the original model.
Equation (\ref{eq:batch}) defines the version of the minority game analysed in this paper.
It has been argued
\cite{GarrMoroSher00} that (\ref{eq:batch}) can be converted into a continuous time
limit of equation (\ref{eq:online}), upon replacing $[q_i(t+1)-q_i(t)]/\sqrt{N}$  by
$dq_i/dt$. Strictly speaking, this is not true. A number of
agents change their preferred strategy at every iteration of
equation (\ref{eq:batch}). The size of their $q$'s will be of the order of
(half) the step size. In the continuous time limit, in contrast,
this step size is lost; yet any discretisation
used to integrate the continuous time differential equation obtained will
effectively re-introduce an (arbitrary) scale for the $q's$. This is not so
relevant when the only appearance of the $q_i$ is in $\sgn[q_i]$, but it is
of importance  in the so-called Thermal Minority game \cite{CavaGarrGiarSher99}, where terms
like $\tanh[\beta q_i]$ appear. We therefore prefer the difference equation
(\ref{eq:batch}) over its continuous counterpart, and regard (\ref{eq:batch}) as the equivalent
of what in the neural networks literature would be called the `batch' version of the
conventional `on-line' minority game. For a more detailed discussion
concerning the validity of a continuous time differential equation for
the TMG we refer to  \cite{CavaGarrGiarSher99,ChalMarsZecc00,MarsChal01}.
Finally, the magnitude of the market fluctuations, or \emph{volatility}, is given by
$\sigma^2=\bra A^2\ket-\bra A\ket^2$. From the starting point
$A(\ell)=N^{-\frac{1}{2}}\sum_i[\omega_i^{\mu(\ell)}+s_i(\ell)\xi_i^{\mu(\ell)}]$
and on the time scales of the process (\ref{eq:batch}), one easily derives
\begin{eqnarray}
\bra A\ket&=&\frac{1}{\alpha N\sqrt{N}}\sum_i
s_i\sum_{\mu}\xi_i^\mu+\order(\frac{1}{\sqrt{N}}),
\label{eq:average}
\\
\label{eq:volatility}
\bra A^2\ket&=&\frac{1}{2}+\frac{1}{\alpha N}[\sum_i h_i s_i
   + \frac{1}{2}\sum_{ij}  s_i J_{ij} s_j]+
   \order(\frac{1}{\sqrt{N}}).
 \end{eqnarray}
 Purely random trading corresponds to $\bra A\ket=0$ and
 $\sigma^2=1$.
We will also define a more general object, the volatility matrix
$\Xi_{tt^\prime}$:
\begin{equation}
\Xi_{tt^\prime}=\bra [A_t-\bra A_t\ket][A_{t^\prime}-\bra A_{t^\prime}\ket]\ket
\label{eq:volatility_matrix}
\end{equation}
which measures the temporal correlations of the market fluctuations.
Note that $\sigma^2_t=\Xi_{tt}$.
In the case where
 the average bid $\langle A \rangle$ is zero (which will turn out to happen in the present
 model), the volatility measures the efficiency of the market.
 Zero volatility
implies that supply and demand are always at the same level, and that
the market is extremely efficient. A large volatility implies large
mismatches between supply and demand, and is the signature of an inefficient
market.

\section{The Generating Functional}

There are two compelling reasons for studying the dynamics of the
minority game. Firstly, dynamical techniques do not rely on the
presence of a Lyapunov-function, so that the MG can be studied for
small $\alpha$. Secondly, it is clear from our
simulations, see the figures below, that, at least on the relevant
time-scales, the stationary state of the minority game can depend
quite strongly on the initial conditions.  One canonical tool to deal
with the dynamics of the present problem is generating functional
analysis \`{a} la De Dominicis \cite{DeDo78}, originally developed in
the disordered systems community (to study spin glasses, in
particular). This formalism allows one to carry out the disorder
average (which here is an average over all strategies) and take the
$N\to\infty$ limit exactly. The final result of the analysis is a set
of closed equations, which can be interpreted as describing the
dynamics of an effective `single agent' \cite{DeDo78,SompZipp82}.  Due
to the disorder in the process, this single agent will acquire an
effective `memory', i.e.  she will evolve according to a non-trivial
non-Markovian stochastic process.

First we rewrite equation (\ref{eq:batch}) as a Chapman-Kolmogorov equation
describing the temporal evolution of an ensemble of markets:
\[
  p_{t+1}(\bq)=\int\!d\bq\, W(\bq|\bq')p_t(\bq'),
\]
where, in the absence of noise, the transition probability density is
simply
\begin{eqnarray*}
  W(\bq|\bq')\!&=&\!
  \prod_i \delta(
     q_i - q_i' +
   h_i + \sum_j J_{ij} s'_j)
  \\
  \!&=&\!
  \int\!\!  \frac{d\hat{\bq}}{(2\pi)^N}
  e^{\sum_i i\hq_i (
     q_i - q_i' +
   h_i + \sum_j J_{ij} s'_j)
  }
\end{eqnarray*}
with the short-hand $s_j^\prime= {\rm sgn}[q_j^\prime]$.
The moment generating functional for a stochastic process of the
present type is defined as
\begin{eqnarray*}
  Z[\bpsi]
  &=&
  \langle~e^{i\sum_t \sum_i  \psi_i(t) q_i(t)}~\rangle
  \\
  &=&
  \int\prod_t\left[ d\bq(t)~ W(\bq(t+1)|\bq(t))\right]~p_0(\bq(0))
  \\
  &&
   \times ~e^{i\sum_t \sum_i  \psi_i(t) q_i(t)}
\end{eqnarray*}
By taking suitable derivatives of the generating functional with respect to the
conjugate variables $\bpsi$, one can generate all moments of $\bq$
at arbitrary times. Upon introducing the two short-hands:
\[
  w^\mu_t=\frac{1}{\tau\sqrt{N}}\sum_i \hq_i(t) \xi^\mu_i, \qquad
  x^\mu_t=\frac{1}{\tau\sqrt{N}}\sum_i s_i(t)\xi^\mu_i,
\]
as well as
$D\bq=\prod_{it}[dq_i(t)/\sqrt{2\pi}]$,
$D\bw=\prod_{\mu t}[dw^\mu_t/\sqrt{2\pi}]$ and
$D\bx=\prod_{\mu t}[dx^\mu_t/\sqrt{2\pi}]$
(with similar definitions for $D\hat{\bq}$, $D\hat{\bw}$ and $D\hat{\bx}$, respectively),
the generating functional takes the following form:
\begin{eqnarray}
  Z[\bpsi]
  &=&
  \int\! D\bw D\hat{\bw} D\bx D\hat{\bx}~
  e^{i\sum_{t \mu}[\hw^\mu_t w^\mu_t +
        \hx^\mu_t x^\mu_t +  w^\mu_t(\Omega^\mu/\tau +
        x^\mu_t)]} \nonumber
  \\&&
  \times \int\! D\bq D\hat{\bq}~p_0(\bq(0))~
   e^{\frac{-i}{\tau\sqrt{N}}\sum_{\mu i}\xi_i^\mu\sum_t [
      \hw^\mu_t \hq_i(t) + \hx^\mu_t s_i(t) ]} \nonumber
  \\&&
  \times ~e^{i\sum_{t i}\left[\hq_i(t) (
     q_i(t+1) - q_i(t)-\theta_i(t))+\psi_i(t) q_i(t)\right]}
 \label{eq:Zbeforeaverage}
\end{eqnarray}
where we have introduced auxiliary driving forces $\theta_i(t)$ to generate
averages involving $\hq_i(t)$ (these can be removed later).

\section{Disorder Averaging}

At this stage we can carry out the disorder averages, to be
denoted as $\overline{\cdots}$, which involve the variables $\xi_i^\mu=\tau^2(R^\mu_{i1}-R^\mu_{i2})$
and $\Omega^\mu=N^{-\frac{1}{2}}\tau^2\sum_j(R^\mu_{j1}+R^\mu_{j2})$
only. For times which do not scale with $N$ one obtains:
\[
\overline{  e^{\frac{i}{\tau}\sum_{t \mu} w^\mu_t \Omega^\mu-\frac{i}{\tau\sqrt{N}}
  \sum_{\mu i}\xi_i^\mu\sum_t [
      \hw^\mu_t \hq_i(t) + \hx^\mu_t s_i(t) ]}}
      ~~~~~~~~~~
\]
\[
=\prod_{i\mu} \overline{e^{\frac{i\tau}{\sqrt{N}}\sum_{t}\left[
w^\mu_t (R_{1}+R_{2})
-(R_{1}-R_{2}) [
      \hw^\mu_t \hq_i(t) + \hx^\mu_t s_i(t) ]\right]}}
\vspace*{-2mm}
\]
\[
=
e^{-\frac{1}{2}\sum_{\mu t t^\prime} \left[
 w^\mu_t w^\mu_{t^\prime}
 +\hw^\mu_t L_{t t^\prime} \hw^\mu_{t^\prime}
 +2 \hx^\mu_{t} K_{t t^\prime} \hw^\mu_{t^\prime}
 +\hx^\mu_t C_{t,t^\prime}\hx^\mu_{t^\prime}
\right] +\order(N^0)}
\]
where we
 have introduced
$C_{t t^\prime}=N^{-1}\sum_i s_i(t) s_i(t^\prime)$,
$K_{t t^\prime}=N^{-1}\sum_i s_i(t) \hq_i(t^\prime)$, and
$L_{t t^\prime}=N^{-1}\sum_i \hq_i(t) \hq_i(t^\prime)$.
We isolate these functions via the insertion of appropriate
$\delta$-functions (in integral representation), and define
the corresponding short-hands $DC=\prod_{t t^\prime}[dC_{t
t^\prime}/\sqrt{2\pi}]$, $DK=\prod_{t t^\prime}[dK_{t
t^\prime}/\sqrt{2\pi}]$ and $DL=\prod_{t t^\prime}[dL_{t
t^\prime}/\sqrt{2\pi}]$
(with similar definitions for $D\hat{C}$, $D\hat{K}$ and $D\hat{L}$,
respectively).
Upon assuming simple initial conditions of the form $p_0(\bq)=\prod_i
p_{0}(q_i)$, the $i$-dependent terms in the disorder-averaged generating functional (\ref{eq:Zbeforeaverage})
are now found to factorise fully over the $N$ traders,
and we arrive at an expression of the following form:
\begin{equation}\label{eq:Zafteraverage}
  \overline{Z[\bpsi]}=
  \int\![DC D\hat{C}][DK D\hat{K}][ DL D\hat{L}]~
    e^{N\left[\Psi+\Phi+\Omega \right]+\order(N^0)}
\end{equation}
The sub-dominant $\order(N^0)$ term in the exponent is
independent of the generating fields $\{\psi_i(t)\}$ and
$\{\theta_i(t)\}$.
There are three distinct leading contributions to the exponent in (\ref{eq:Zafteraverage}).
The first is a
`bookkeeping' term, linking the two-time order parameters to their
conjugates:
\[
  \Psi=i~ \Tr~ [ \hC^T C + \hK^T K + \hL^T L],
\]
The second reflects the statistical properties of the players'
arsenal of strategies:
\begin{eqnarray}
\Phi
&=&
\alpha  \log\left[ \int\! Dw D\hat{w} Dx D\hat{x}~
  e^{i\sum_{t}[\hw_t w_t + \hx_t x_t +  w_t x_t]}
  \right.
\nonumber\\
&&\times
\left.
e^{-\frac{1}{2}\sum_{t t^\prime} \left[
 w_t w_{t^\prime}
 +\hw_t L_{t t^\prime} \hw_{t^\prime}
 +2\hx_{t} K_{t t^\prime} \hw_{t^\prime}
 +\hx_t C_{t t^\prime}\hx_{t^\prime}
\right]}\right]
\label{eq:Phi}
\end{eqnarray}
The third term, which contains the generating fields, will describe the (now
stochastic) evolution of the strategy valuations $q(t)$ of a single effective agent:
\begin{eqnarray*}
\Omega &=& \frac{1}{N}\!\sum_i\log
\left[ \int\! Dq D\hat{q}~p_0(q(0))~
e^{i\sum_{t}\hq(t)[
     q(t+1) - q(t)-\theta_i(t)]}
\nonumber\right.\\
&&
\left.\hspace*{-5mm}
\times~
e^{i\sum_{t}\psi_i(t) q(t)-i\sum_{t t^\prime}[s(t) \hat{C}_{t t^\prime}s(t^\prime)
+s(t)\hat{K}_{t t^\prime}  \hq(t^\prime)
+\hq(t)\hat{L}_{t t^\prime}  \hq(t^\prime)]}
     \right]
\end{eqnarray*}
with $s(t)={\rm sgn}[q(t)]$ and with
$Dq=\prod_{t}[dq(t)/\sqrt{2\pi}]$,
$Dw=\prod_{t}[dw_t/\sqrt{2\pi}]$,
$Dx=\prod_{t}[dx_t/\sqrt{2\pi}]$
(similar definitions for $D\hat{q}$, $D\hat{w}$ and
$D\hat{x}$).
The form of Eq. (\ref{eq:Zafteraverage}) is suitable for a saddle-point integration
in the thermodynamic limit $N\to\infty$.
With a modest amount of foresight we define $G_{t t^\prime}=-iK_{t t^\prime}$. Upon taking derivatives with respect to the generating fields
$\{\theta_i(t),\psi_i(t)\}$, and using the built-in normalisation $\overline{Z[{\bf
0}]}=1$, we find that {\em at} the relevant saddle-point:
\begin{eqnarray}
C_{t t^\prime}&=&\lim_{N\to\infty}\frac{1}{N}\sum_i\overline{\bra
s_i(t) s_i(t^\prime)\ket}
\label{eq:meaningof_C}
\\
G_{t t^\prime}&=&\lim_{N\to\infty}\frac{1}{N}\sum_i\frac{\partial}{\partial
\theta_i(t^\prime)}\overline{\bra s_i(t)\ket}
\label{eq:meaningof_K}
\\
L_{t t^\prime}&=&
\lim_{N\to\infty}\frac{1}{N}\sum_i\frac{\partial^2}{\partial
\theta_i(t^\prime)\partial\theta_i(t^\prime)}\overline{Z[{\bf 0}]}=0
\label{eq:meaningof_L}
\end{eqnarray}
The first two are recognised to represent
disorder-averaged and site-averaged correlation- and response
functions.
At this stage the generating fields are in principle no longer needed. We will
put $\psi_i(t)=0$ and $\theta_i(t)=\theta(t)$, and find our
expression for $\Omega$ simplifying to
\begin{eqnarray}
\Omega &=&\log
\left[ \int\! Dq D\hat{q}~p_0(q(0))~
e^{i\sum_{t}\hq(t)[
     q(t+1) - q(t)-\theta(t)]}
\nonumber\right.\\
&&
\left.
\times~
e^{-i\sum_{t t^\prime}[s(t) \hat{C}_{t t^\prime}s(t^\prime)
+s(t)\hat{K}_{t t^\prime}  \hq(t^\prime)
+\hq(t)\hat{L}_{t t^\prime}  \hq(t^\prime)]}
     \right]
\label{eq:Omega}
\end{eqnarray}
Extremisation of the extensive exponent $\Psi+\Phi+\Omega$ of (\ref{eq:Zafteraverage}) with respect to
$\{C,\hat{C},K,\hat{K},L,\hat{L}\}$ gives the saddle-point equations
\begin{eqnarray}
C_{t t^\prime}=\langle s(t) s(t^\prime) \rangle_\star
~~~~~~~~
G_{t t^\prime}=\frac{\partial \bra
s(t)\ket_\star}{\partial\theta({t^\prime})}
~~~~~~~~
\label{eq:CandG}
\\
\hC_{tt^\prime}=\frac{i\partial \Phi}{\partial C_{t t^\prime}}
~~~~~~
\hK_{tt^\prime}=\frac{i\partial \Phi}{\partial K_{t t^\prime}}
~~~~~~
\hL_{tt^\prime}=\frac{i\partial \Phi}{\partial L_{t t^\prime}}
\label{eq:Conjugates}
\end{eqnarray}
whereas $L_{t t^\prime}=0$. The effective single-trader
averages $\bra \ldots\ket_\star$, generated by taking
derivatives of (\ref{eq:Omega}), are defined as follows (note: $s(t)={\rm sgn}[q(t)]$):
\[
\bra f[\{q\}]\ket_\star= \frac{\int\! Dq ~M[\{q\}]f[\{q\}]}{\int\! Dq
~M[\{q\}]}
\]
\begin{eqnarray}
M[\{q\}]&=& p_0(q(0))e^{-i\sum_{t t^\prime} s(t) \hat{C}_{t
t^\prime}s(t^\prime)}
\nonumber \\
&&\times
\int\! D\hat{q}~e^{-i\sum_{t t^\prime}\hq(t)\hat{L}_{t t^\prime}\hq(t^\prime)}
\nonumber \\
&&\times~
e^{i\sum_{t}\hq(t)[q(t+1) - q(t)-\theta(t)-\sum_{t^\prime}\hat{K}^T_{t t^\prime}s(t^\prime)]}
\label{eq:singletrader_measure}
\end{eqnarray}
Upon elimination of $\{\hat{C},\hat{K},\hat{L}\}$ via (\ref{eq:Conjugates}),
we have now obtained exact closed equations
for the disorder-averaged correlation- and response functions in the $N\to\infty$ limit: namely
(\ref{eq:CandG}), with the effective single trader measure
(\ref{eq:singletrader_measure}).

\section{Simplification of the Saddle-Point Equations}

The above procedure is quite insensitive to changing model details;
alternative choices made for the statistics of traders' strategies would
simply lead to a different form for the function $\Phi$
(\ref{eq:Phi}), whereas changing the update rules for the strategy valuations of
the  traders (e.g. by making these non-deterministic, as in
\cite{CavaGarrGiarSher99,ChalMarsZecc00}) would only affect the details of the term $\Omega$
(\ref{eq:Omega}). We now work out our equations for the present
choice of model.
Focusing first on $\Phi$, we perform the $x_t$ integrals, yielding
$\prod_t\delta[\hx_t+w_t]$, and after performing the remaining $\hx$
integrations we get
\begin{eqnarray*}
\Phi&=&
\alpha  \log \int\! Dw D\hat{w}~
  e^{i\sum_{t}\hw_t w_t}
\\
&&e^{-\frac{1}{2}\sum_{t t^\prime} [
 w_t w_{t^\prime}
 +\hw_t L_{t t^\prime} \hw_{t^\prime}
 -2w_{t} K_{t t^\prime} \hw_{t^\prime}
 +w_t C_{t t^\prime}w_{t^\prime}]}
\end{eqnarray*}
The Gaussian integration over the $\{w_t\}$ gives
\begin{eqnarray*}
  \Phi&=&-\frac{1}{2}\alpha \log \det D
  +\alpha \log \int\!\prod_t\left[\frac{d\hat{w}_t}{\sqrt{2\pi}}\right]
  e^{-\frac{1}{2}\sum_{tt^\prime}\hat{w}_t
  L_{tt^\prime}\hat{w}_{t^\prime}}
\nonumber \\
&& \times~e^{-\frac{1}{2}\sum_{tt^\prime}\hw_t\left[
(\openone-i K)^T
D^{-1}(\openone-iK)\right]_{tt^\prime}\hw_{t^\prime}}
\end{eqnarray*}
where the entries of the matrix $D$ are given by $D_{tt^\prime}= 1+C_{tt^\prime}$.
We now take the derivative of $\Phi$ with respect to $L_{tt^\prime}$, as
dictated by (\ref{eq:Conjugates}), and subsequently put all $L_{tt^\prime}\to 0$.
This gives
\[
  \hL=-\frac{1}{2}i\alpha
    (\openone-i K)^{-1}
    D(\openone-i K^T)^{-1},
\]
and $\lim_{L\to 0}\Phi= -\alpha\Tr\log(\openone-iK)$, so that
\[
  \hK^T=-\alpha(\openone-i K)^{-1}~~~~~~~~~~~\hC=0
\]
We now write our final result in terms of the response function
(\ref{eq:meaningof_K}), via the identity $K=iG$, and find
our effective single trader measure $M[\{q\}]$ of
(\ref{eq:singletrader_measure}) reducing to
\begin{eqnarray}
p_0(q(0))
\int\! D\hat{q}~e^{-\frac{1}{2}\alpha\sum_{t t^\prime}\hq(t)[(\openone+G)^{-1} D(\openone+G^T)^{-1}]_{t t^\prime}
\hq(t^\prime)}
\nonumber \\
\times~
e^{i\sum_{t}\hq(t)[q(t+1) - q(t)-\theta(t)+\alpha\sum_{t^\prime}(\openone+G)^{-1}_{t t^\prime}
s(t^\prime)]}
\label{eq:singleagent_statistics}
\end{eqnarray}
This describes a stochastic single-agent process of the form
\begin{equation}
\label{eq:singleagent}
  q(t\!+\!1) = q(t)+\theta(t) - \alpha \sum_{t^\prime\leq t} (\openone+ G)^{-1}_{t t^\prime} {\rm sgn}[q(t^\prime)]
  +\sqrt{\alpha}
\eta(t),
\end{equation}
Causality ensures that $G_{tt^\prime}=0$ for all $t^\prime\geq t$
(so that $(\openone+ G)^{-1}_{tt^\prime}=0$ for $t^\prime >t$), and
 $\eta(t)$ is a Gaussian noise with zero mean and with
  temporal correlations given by
  $\bra\eta(t)\eta(t^\prime)\ket=\Sigma_{tt^\prime}$:
\begin{equation}
\Sigma =
(\openone+G)^{-1} D(\openone+G^T)^{-1}
\label{eq:noise_covariance}
\end{equation}
The correlation- and response functions, defined by
(\ref{eq:meaningof_C},\ref{eq:meaningof_K}), are the dynamic
order parameters of the problem, and must be solved
self-consistently from the closed equations
\begin{equation}
C_{t t^\prime}=\langle {\rm sgn}[q(t) q(t^\prime)] \rangle_\star
~~~~~~~~
G_{t t^\prime}=\frac{\partial \bra {\rm sgn}[q(t)]\ket_\star}{\partial\theta({t^\prime})}
\label{eq:finalCandG}
\end{equation}
Note that $M[\{q\}]$ as given by
(\ref{eq:singleagent_statistics})
is normalised, i.e. $\int\! Dq~M[\{q\}]=1$, so
 the associated averages reduce to $\bra f[\{q\}]\ket_\star= \int\! Dq
~M[\{q\}]f[\{q\}]$.
The solution of (\ref{eq:finalCandG})
can be calculated numerically with arbitrary precision, without
finite size
effects, using a technique described in \cite{EissOppe92}.

Finally, in appendix \ref{sec:average_and_volatility} we
calculate the disorder averaged re-scaled average bid $\overline{\bra A_t\ket}$ and
 volatility matrix $\overline{\Xi}_{tt^\prime}=
 \overline{\bra A_t A_{t^\prime}\ket}-\overline{\bra A_t\ket \bra
 A_{t^\prime}\ket}$,
for $N\to\infty$,
as defined previously in (\ref{eq:average}) and
(\ref{eq:volatility_matrix}).
Note that objects such as $\overline{\bra A_t\ket}$ must asymptotically become self-averaging,
i.e. independent of the microscopic realisation of the
disorder; hence $\overline{\bra A_t\ket \bra A_{t^\prime}\ket}\to \overline{\bra A_t\ket}
~\overline{\bra A_{t^\prime}\ket}$ for $N\to\infty$.
We find the satisfactory result that the average bid is zero, and that the volatility
matrix (and thus also the ordinary single-time volatility
$\sigma_t^2=\Xi_{tt}$)
is proportional to the covariance matrix
(\ref{eq:noise_covariance}) of the noise in the dynamics
(\ref{eq:singleagent}) of the effective single agent:
\begin{equation}
\lim_{N\to\infty}\overline{\bra A\ket_t}=0,~~~~~~~~
\lim_{N\to\infty}\overline{\Xi}_{tt^\prime}=\frac{1}{2}\Sigma_{t t^\prime}
\label{eq:results_of_appendix}
\end{equation}
Thus the noise term $\eta(t)$ in the single agent process
(\ref{eq:singleagent}) represents the overall market
fluctuations, and the covariance matrix
(\ref{eq:noise_covariance}) informs us of both single-time volatility and
the temporal correlations of the market fluctuations.

\section{The First Time Steps} \label{sec:first}

For the first few time steps it is possible to calculate
the order parameters (correlation- and response functions) and the volatility
explicitly, starting from the effective single trader measure
(\ref{eq:singleagent_statistics}). Note that $D_{tt^\prime}=1+C_{tt^\prime}$ and that
$C_{tt}=1$ for any $t$. Significant
simplifications can be made by using causality. For instance, we
always have $(\openone+G)^{-1}=\sum_{n\geq 0}(-1)^n G^n$, with
causality enforcing
\begin{equation}
[G^{n}]_{tt^\prime}=0~~~{\rm for}~~~t^\prime>t-n
\label{eq:causality}
\end{equation}
At $t=0$ this immediately allows us to conclude that
$\Sigma_{00}=D_{00}=2$.
We now obtain from (\ref{eq:singleagent_statistics}) the joint statistics at times $t=1$:
\begin{equation}
p(q(1)|q(0))=\frac{e^{-\left[q(1)-q(0)-\theta(0)+\alpha~ {\rm
sgn}[q(0)]\right]^2/4\alpha}}{2\sqrt{\alpha \pi}}
\label{eq:joint_dist_t=01}
\end{equation}
Equation (\ref{eq:joint_dist_t=01}), in turn, allows us to calculate $C_{10}=\bra\sgn[q(0)q(1)]\ket_\star$
and $G_{10}=\partial \bra\sgn[q(1)]\ket_\star/\partial \theta(0)$:
\begin{eqnarray*}
C_{10}&=&-\int\!dq(0)~p(q(0))~\erf\left[\frac{\sqrt{\alpha}}{2}-\frac{|q(0)|\!+\theta(0){\rm
sgn}[q(0)]}{2\sqrt{\alpha}}\right]
\\
G_{10}
&=&\frac{1}{\sqrt{\alpha \pi}}
\int\!dq(0)~p(q(0))~
e^{-\left[\alpha~{\rm sgn}[q(0)]-q(0)-\theta(0)\right]^2/4\alpha }
\end{eqnarray*}
We can now move to the next time step, again using (\ref{eq:causality}), where we need the noise
covariances $\Sigma_{11}$ and $\Sigma_{10}$:
\begin{eqnarray*}
\Sigma_{10}&=&\sum_{tt^\prime}[\openone-G+\order(G^2)]_{1t}D_{tt^\prime}
[\openone-G^T+\order(G^{T})^2]_{t^\prime 0}
\\
&=& 1+C_{10}-2G_{10}
\\[2mm]
\Sigma_{11}&=&\sum_{tt^\prime}
[\openone-G+\order(G^2)]_{1t} D_{tt^\prime}[\openone-G^T+\order(G^{T})^2]_{t^\prime 1}
\\
&=& 2-2G_{10}[1+ C_{01}]
+2[G_{10}]^2
\end{eqnarray*}
Although this procedure can in principle be repeated for an arbitrary number of time steps,
generating exact expressions for the various order parameters iteratively,
the results become increasingly complicated when larger times are involved.

It is interesting, however, to inspect further some special limits.
We first turn to the (trivial) case where $\alpha$
is very small, $p(q(0))=\delta[q(0)-q_0]$ and $q_0$ is finite.
Provided $|q_0|\gg \sqrt{\alpha}$ as $\alpha\to 0$, we immediately deduce
from the above results that $\lim_{\alpha\to 0}C_{10}=1$, $\lim_{\alpha\to
0}G_{10}=0$, and $\lim_{\alpha\to 0}\Sigma_{10}=\lim_{\alpha\to
0}\Sigma_{11}=2$. Hence we find in leading order in $\alpha$ that
$q(1)=q(0)$ and $\eta(1)=\eta(0)$. One easily repeats the argument
for larger times, and finds that, without perturbations, both the system variables $q(t)$ and the noise variables
$\eta(t)$  will remain frozen for times $t\ll
1/\sqrt{\alpha}$, the only remaining uncertainty in the noise being the realisation of
$\eta(0)$:
\[
q(t)=q_0+t\sqrt{\alpha}~\eta(0)+\order(\alpha t)
~~~~~~~~(\alpha\to 0)
\]
If $\sgn[q_0]\neq \sgn[\eta(0)]$,
the system will `de-freeze'
at the first instance where $t>|q_0/\eta(0)\sqrt{\alpha}|$.
 Since $\eta(0)$ is a zero average Gaussian
variable, one should therefore for small $\alpha$ expect half of the population of traders
(those with non-profitable initial random strategy choices) to
commence strategy chances at time-scales
$t=\order(\alpha^{-\frac{1}{2}})$,
whereas the other half will continue playing the game
with their (for now profitable) initial strategy choices at least up to
$t=\order(\alpha^{-1})$.

 It is also interesting to analyse the case where the game  is initialised
in a \emph{tabula rasa} manner (which appears to have been common practice in literature),
i.e. $p(q(0))=\delta[q-q_0]$ with
$q_0=0^{+}$, and where we have no perturbation fields, i.e. $\theta(t)=0$.
Now the above results reduce to
\begin{eqnarray*}
C_{10}&=&-\erf[\frac{1}{2}\sqrt{\alpha}]
~~~~~~~~~~
G_{10}=(\alpha \pi)^{-\frac{1}{2}} e^{-\frac{1}{4}\alpha}
\\
\Sigma_{10}&=& 1-\erf[\frac{1}{2}\sqrt{\alpha}]-\frac{2}{\sqrt{\alpha \pi}} e^{-\frac{1}{4}\alpha}
\\
\Sigma_{11}&=& 2-\frac{2}{\sqrt{\alpha \pi}} e^{-\frac{1}{4}\alpha}[1-\erf[\frac{1}{2}\sqrt{\alpha}]]
+\frac{2}{\alpha \pi} e^{-\frac{1}{2}\alpha}
\end{eqnarray*}
The negative value of the correlation function $C_{10}$ implies that for short times
the traders will exhibit a tendency to alternate their (two)
strategies.
Let us now inspect the limiting behaviour of the above expressions
for large and small values of $\alpha$. For large $\alpha$ one
obtains
\[
\lim_{\alpha\to\infty}C_{10}=\lim_{\alpha\to\infty}G_{10}=\lim_{\alpha\to\infty}\Sigma_{10}=0
\]
In other words, the agents trade independently and randomly; for
larger times this will continue to be the case. For small
$\alpha$, on the other hand, we find
\begin{eqnarray*}
C_{10}&=&-\frac{\sqrt{\alpha}}{\sqrt{\pi}}+\order(\alpha^{\frac{3}{2}})
~~~~~~~~~~~
G_{10}=\frac{1}{\sqrt{\alpha \pi}}+\order(\sqrt{\alpha})
\\
\Sigma_{10}&=&
1-\frac{2}{\sqrt{\alpha \pi}}+\order(\sqrt{\alpha})
~~~~~~
\Sigma_{11}=\frac{2}{\alpha \pi}-\frac{2}{\sqrt{\alpha \pi}}+\order(\alpha^0)
\end{eqnarray*}
So $\eta(1)=\order(\alpha^{-\frac{1}{2}})$, whereas
$\eta(0)=\order(\alpha^0)$. We also find
\[
\bra [\eta(1)+\frac{\eta(0)}{\sqrt{\alpha\pi}}]^2\ket=
\Sigma_{11}+\frac{2}{\sqrt{\alpha\pi}}\Sigma_{10}+\frac{1}{\alpha
\pi}\Sigma_{00}=\order(\alpha^0)
\]
from which it follows that $\eta(1)=-\eta(0)/\sqrt{\alpha\pi}+\order(\alpha^0)$, and hence
we can write the first steps of the effective single agent equation (\ref{eq:singleagent}) as
\begin{eqnarray*}
q(1) &=& q(0)- \alpha~ {\rm sgn}[q(0)] +\sqrt{\alpha}\eta(0)
\\
&=& \sqrt{\alpha}\eta(0)+\order(\alpha)
\\
q(2) &=& q(1) -\alpha ~{\rm sgn}[q(1)]+\alpha G_{10}{\rm sgn}[q(0)]+\sqrt{\alpha}\eta(1)
\\
&=& -\eta(0)/\sqrt{\pi}+\order(\sqrt{\alpha})
\end{eqnarray*}
Thus also $C_{20}=\bra \sgn[q(0)q(2)]\ket_\star=\order(\sqrt{\alpha})$
and $C_{21}=\bra
\sgn[q(1)q(2)]\ket_\star=-1+\order(\sqrt{\alpha})$.
We observe that for small $\alpha$ the first two time steps are
driven predominantly by the noise component in
(\ref{eq:singleagent}). This noise component increases in strength
and starts oscillating in sign, resulting in an effective agent which
is increasingly likely to alternate its strategies.
Equivalently, this implies that in the initial $N$-agent system an increasing
{\em fraction} of the population of agents will start alternating their strategies.

Let us finally inspect the initial behaviour of equation
(\ref{eq:singleagent}) for the intermediate regime where $p(q(0))=\delta[q-q_0]$
with $q_0=\order(\sqrt{\alpha})$, to which  (as we have seen) also for
$q_0=\order(\alpha^0)$ about half of the traders
will automatically  be driven in due course.
We now put $q_0=\sqrt{\alpha}\tilde{q}_0$ and find in leading
order:
\begin{eqnarray*}
C_{10}&=&\erf[\frac{1}{2}|\tilde{q}_0|]+\ldots
~~~~~~~~~~~
G_{10}=\frac{1}{\sqrt{\alpha \pi}}
e^{-\frac{1}{4}\tilde{q}^2_0}+\ldots
\\
\Sigma_{10}&=& -\frac{2}{\sqrt{\alpha \pi}}
e^{-\frac{1}{4}\tilde{q}^2_0}+\ldots
~~~~~~
\Sigma_{11}=
\frac{2}{\alpha \pi}
e^{-\frac{1}{2}\tilde{q}^2_0}+\ldots
\end{eqnarray*}
Thus we have $\bra[\eta(1)+(\alpha
\pi)^{-\frac{1}{2}}e^{-\tilde{q}^2_0/4}\eta(0)]^2\ket=0$,
so also $\eta(1)=-(\alpha
\pi)^{-\frac{1}{2}}e^{-\tilde{q}^2_0/4}\eta(0)$,
in leading order for $\alpha\to 0$.
This then, together with $q(1)=\order(\sqrt{\alpha})$ (which
immediately follows from (\ref{eq:joint_dist_t=01})), leads us to
\[
q(2)= -\pi^{-\frac{1}{2}}e^{-\frac{1}{4}\tilde{q}^2_0}\eta(0)+
  \order(\sqrt{\alpha})
\]
We thus find that also for $q_0=\order(\sqrt{\alpha})$ the initial conditions are
more or less washed out by the internal noise generated by the process,
within just two iteration steps.


\section{The Stationary State for $\alpha>\alpha_c$}

For general $\alpha$, not necessarily small, the arguments used in
the second part of the previous section do not hold. In a
stationary state, along with agents that will change strategy
(almost) every cycle, there will generally also be agents finding
themselves consistently in the minority group, which will
consequently play the same strategy over and over again. For the
latter `frozen' group (a term introduced in \cite{ChalMars99}), 
the differences between the valuations of
the two available strategies (i.e. the values of $q_i$) will grow more
or less linearly in time, whereas the `fickle' agents will have
values for $q_i$ very close to zero. In order to separate the two
groups efficiently we introduce the re-scaled values
$\tilde{q}_i(t)=q_i(t)/t$. Frozen agents will be those for which
$\lim_{t\to\infty}\tilde{q}_i(t)\neq 0$. Similarly the effective
single agent process (\ref{eq:singleagent}) is transformed via
$\tilde{q}(t)=q(t)/t$, where now the quantity $\phi=\lim_{\epsilon\to
0}\lim_{t\to\infty}\bra \theta[|\tilde{q}(t)|-\epsilon]\ket_\star$
will give the asymptotic fraction of `frozen' agents in the
original $N$-agent system, for $N\to \infty$. The dynamical
equation of the re-scaled effective agent can be written as
\begin{eqnarray}
\widetilde{q}(t)&=&
  \frac{1}{t}\widetilde{q}(1)
 + \frac{\sqrt{\alpha}}{t}\sum_{t^\prime <t}\eta(t^\prime)
 \nonumber\\
&& -\frac{\alpha}{t}\sum_{t^\prime <t}
   \sum_{t^\pprime}(\openone+G)^{-1}_{t^\prime t^\pprime}\sgn[\tilde{q}(t^\pprime)]
\label{eq:rescaled_agent}
\end{eqnarray}
If the game has reached a stationary state, then
$G_{tt^\prime}=G(t-t^\prime)$,
$C_{tt^\prime}=C(t-t^\prime)$ and $\Sigma_{t t^\prime}=\Sigma(t-t^\prime)$, by
definition. We will in this section assume that the stationary
\begin{figure}[t]
\centerline{
\epsfig{file=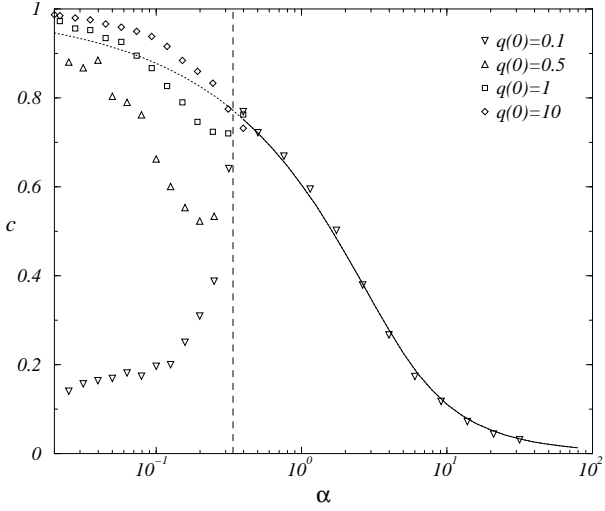, width=80mm}
}
\vspace*{3mm}
\caption{\label{fig:c}
Asymptotic average $c=\lim_{\tau\to\infty}\tau^{-1}\sum_{t\leq \tau}C(\tau)$
of the stationary covariance. The
markers are obtained from individual simulation runs performed with a system of $N=4000$
agents and various
homogeneous initial valuations (where $q_i(0)=q(0)$), and in excess of 1000 iteration steps.
The solid curve to the right of the
critical point is the theoretical prediction, given by the solution of
 (\ref{eq:c}).
The dotted curve to the left is its continuation into the $\alpha<\alpha_c$ regime
(where it should no longer be correct).
 }
\end{figure}
\noindent
state
is one without anomalous response, i.e. temporary perturbations will
not influence the stationary state and
decay sufficiently fast, such that
$\lim_{\tau\rightarrow\infty}\sum_{t\leq \tau} G(t)=k$ exists. This condition will be met if
there is just one ergodic component; it is the dynamical equivalent of replica symmetry
being stable (see e.g. \cite{MezaPariVira87}) in a detailed balance model.
We now define $\tilde{q}=\lim_{t\to\infty}\tilde{q}(t)$
(assuming this limit exists) and take the limit $t\to\infty$ in Eqn.
(\ref{eq:rescaled_agent}). Under the assumption of absent anomalous response,
we can use the two lemmas in appendix B to simplify the result to
\begin{equation}
  \widetilde{q}=-\frac{\alpha}{1+k}s+\sqrt{\alpha}\eta
\label{eq:stationarity}
\end{equation}
with the averages $s=\lim_{\tau\to\infty}\tau^{-1}\sum_{t\leq
  \tau}\sgn[\tilde{q}_t]$ and
  $\eta=\lim_{\tau\to\infty}\tau^{-1}\sum_{t\leq
  \tau}\eta(t)$.
The variance of the zero-average Gaussian random variable $\eta$
follows from
(\ref{eq:noise_covariance}):
\begin{eqnarray}
\bra \eta^2 \ket &=&
\lim_{\tau,\tau^\prime\to\infty}\frac{1}{\tau \tau^\prime}
\sum_{t\leq \tau}\sum_{t^\prime\leq
  \tau^\prime}[(\openone+G)^{-1} D(\openone+G^T)^{-1}]_{t t^\prime}
\nonumber \\
&=&
(1+k)^{-2}[1+\lim_{\tau,\tau^\prime\to\infty}\frac{1}{\tau \tau^\prime}
\sum_{t\leq \tau}\sum_{t^\prime\leq
  \tau^\prime}C_{tt^\prime}]
\nonumber \\
&=& (1+k)^{-2}[1+\bra s^2\ket]
\label{eq:persistent_eta}
\end{eqnarray}
Note that $\bra s^2\ket=\lim_{\tau\to\infty}\tau^{-1}\sum_{t\leq
\tau}C(t)=c$.

The effective agent is frozen if $\widetilde{q}\not=0$, in which case
$s= \sgn[ \widetilde{q}]$. This solves equation (\ref{eq:stationarity}) if and
only if $|\eta|>\sqrt{\alpha}/(1+k)$.  If
$|\eta|<\sqrt{\alpha}/(1+k)$, on the other hand, the
 agent is not frozen; now
$\widetilde{q}=0$ and
$s=(1+k)\eta/\sqrt{\alpha}$.

\begin{figure}[t]
\centerline{
\epsfig{file=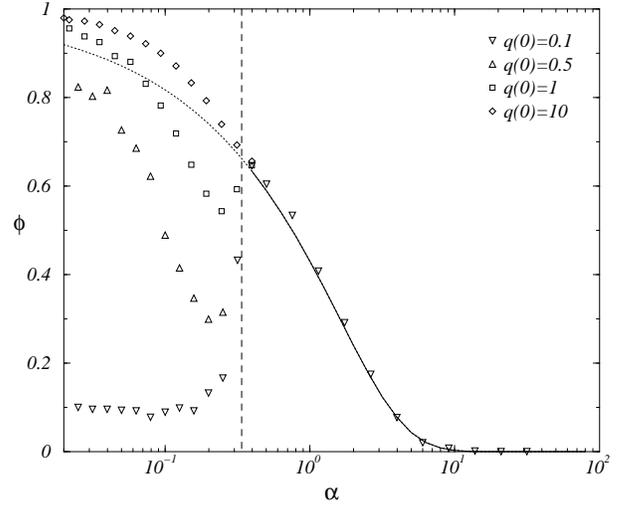,width=80mm}
}
\vspace*{3mm}
\caption{\label{fig:frozen}
Fraction $\phi=1-\erf[\sqrt{\alpha/2(1+c)}]$ of frozen agents
in the stationary state.
The markers are obtained from individual simulation runs performed with a
system of $N=4000$ agents and various
homogeneous initial conditions, where $q_i(0)=q(0)$, and
in excess of 1000 iteration steps.
 The solid line to the right of
the critical point  is the theoretical prediction, obtained from
the solution of (\ref{eq:c}).
The dotted curve to the left is its continuation into the $\alpha<\alpha_c$ regime
(where it should no longer be correct).
}
\end{figure}
\noindent
We can now calculate $c=\bra s^2\ket$ self-consistently, upon distinguishing
between the two possibilities:
\[
c=
\bra\theta\left[|\eta|-\frac{\sqrt{\alpha}}{1+k}\right]\ket
+ \bra\theta\left[\frac{\sqrt{\alpha}}{1+k}-|\eta|\right]
\frac{(1+k)^2\eta^2}{\alpha}\ket
\]
Working out the Gaussian integrals describing the statics of
$\eta$,
with variance (\ref{eq:persistent_eta}), then gives
\begin{equation}
  c=1-(1-\frac{1+c}{\alpha})~\erf\left[\sqrt{\frac{\alpha}{2(1+c)}}\right]
 -2\sqrt{\frac{1+c}{2\pi \alpha}}e^{-\frac{\alpha}{2(1+c)}}.
  \label{eq:c}
\end{equation}
From this equation the value of $c$ can be solved numerically.
For large $\alpha$ the solution behaves
as $c\sim\alpha^{-1}$.
In figures \ref{fig:c}
and \ref{fig:frozen}
we show the solution of
(\ref{eq:c}) and the fraction
$\phi$ of frozen agents, given according to the theory by
$\phi=\bra\theta[
|\eta|-\sqrt{\alpha}/(1+k)]\ket=1-\erf[\sqrt{\alpha/2(1+c)}]$,
as functions of $\alpha$, together with the
values for $c$ and $\phi$ as obtained by carrying out numerical simulations
of the minority game.
One observes excellent agreement between theory and experiment
above a critical value $\alpha_c$, which we will calculate below.

From the time-averaged asymptotic correlation $c$ we next move on
to calculate the integrated response
$k=\lim_{\tau\rightarrow\infty}\sum_{t\leq \tau} G(t)$. Since the occurrence of
the Gaussian noise term $\eta(t)$ in Eqn.
(\ref{eq:singleagent}) is (apart from a factor $\sqrt{\alpha}$) similar to that of
an external field, we can write the response function as
$G_{tt^\prime}=\alpha^{-\frac{1}{2}}\bra \partial~\sgn[q(t)]/\partial \eta(t^\prime)
\ket_\star$. Integration by parts in this expression generates
\[
\bra \partial~\sgn[q(t)]/\partial \eta(t^\prime)
\ket_\star
=\sum_{t^\pprime}\Sigma^{-1}_{t^\prime t^\pprime}
\bra \sgn[q(t)] \eta(t^\pprime)\ket_\star
\]
and hence
\begin{equation}
\sqrt{\alpha}\sum_{t^\pprime}\bra \eta(t)\eta(t^\pprime)\ket
G^T_{t^\pprime t^\prime}
=
\bra \sgn[q(t)] \eta(t^\prime)\ket_\star
\label{eq:noise_relation}
\end{equation}
Averaging over the two times $t$ and $t^\prime$ now gives
in a stationary state, upon using again the assumption of absent
anomalous response
(and the familiar notational conventions $s=\lim_{\tau\to\infty}\tau^{-1}\sum_{t\leq
\tau}\sgn[q(t)]$ and $\eta=\lim_{\tau\to\infty}\tau^{-1}\sum_{t\leq
\tau}\sgn[q(t)]$):
\begin{eqnarray}
\bra s \eta\ket&=&
\sqrt{\alpha}
\lim_{\tau\to\infty}\frac{1}{\tau}\sum_{t^\prime\leq \tau}
\sum_{t^\pprime}\bra \eta \eta(t^\pprime)\ket
G^T_{t^\pprime t^\prime}
\nonumber \\
&=&
k\sqrt{\alpha}\bra \eta^2 \ket
\label{eq:eqn_for_k}
\end{eqnarray}
The variance $\bra \eta^2\ket$ is given in
(\ref{eq:persistent_eta}). We calculate the remaining object
$\bra s \eta\ket$ similarly to our calculation of $c$, by
distinguishing between frozen and non-frozen agents and by using
the two identities $s=\sgn[\eta]$ (for frozen agents) and
$s=\eta(1+k)/\sqrt{\alpha}$ (for the non-frozen ones), both of
which follow immediately from (\ref{eq:stationarity}).
This results in
\begin{eqnarray*}
\bra s \eta\ket&=&
\bra\theta\left[|\eta|-\frac{\sqrt{\alpha}}{1+k}\right]|\eta|\ket
+ \bra\theta\left[\frac{\sqrt{\alpha}}{1+k}-|\eta|\right]
\frac{\eta^2(1+k)}{\sqrt{\alpha}}
\ket
\\
&=&
\frac{1+c}{(1+k)\sqrt{\alpha}}~
\erf\left[\sqrt{\frac{\alpha}{2(1+c)}}\right]
\end{eqnarray*}
Insertion into (\ref{eq:eqn_for_k}), together with
(\ref{eq:persistent_eta}), then gives the desired expression for
the integrated response:
\begin{equation}\label{eq:k}
  \frac{1}{k}=\frac{\alpha}{\erf[\sqrt{\frac{\alpha}{2(1+c)}}]}-1
\end{equation}
with the value of $c$ to be determined by solving Eqn.
(\ref{eq:c}). Equivalently,
using $\phi=1-\erf[\sqrt{\alpha/2(1+c)}]$ we get
\begin{equation}
k=\frac{1-\phi}{\alpha-1+\phi}
\label{eq:k2}
\end{equation}
The integrated response $k$ is positive and finite, and hence our solution
(based on this property) is exact, for $\alpha>\alpha_c$. Here $\alpha_c$
is the point at which $k$ diverges, which is found to happen when
the fraction of fickle agents equals $\alpha$.  According to
(\ref{eq:c},\ref{eq:k}) we can write $\alpha_c$ as $\alpha_c=\erf[x]$, where $x$
is the solution of the transcendental equation
\begin{equation}
 \erf[x]=2 -\frac{1}{x\sqrt{\pi}}e^{-x^2}
\end{equation}
The resulting value
$\alpha_c\approx 0.33740$ is identical to that found in $\cite{ChalMarsZecc00}$
(for a stochastic version of the game) using replica calculations. Below $\alpha_c$
there might well be multiple ergodic components,
i.e. more than one stationary solution of our  fundamental order
parameter equations (\ref{eq:finalCandG}).

\section{Stationary Volatility for $\alpha>\alpha_c$}

In contrast to the persistent order parameter $c$ and its
relative $k$, the volatility matrix (\ref{eq:volatility_matrix}), to be calculated within
our theory from expressions (\ref{eq:noise_covariance},\ref{eq:results_of_appendix})
and in a stationary state of the Toeplitz form
$\Xi_{tt^\prime}=\Xi(t-t^\prime)$,
generally involves both
long-term and short-term fluctuations.
This becomes apparent when we work out $\Xi(t)$ using
(\ref{eq:noise_covariance}) and the results of
appendix B.
We separate in the functions $C$ and $G$ the persistent from the
non-persistent terms, i.e. $C(t)=c+\tilde{C}(t)$ and
$G(t)=\tilde{G}(t)$ (there is no persistent response for $\alpha>\alpha_c$),
and find
\begin{eqnarray}
2\Xi(t)
&=& \frac{1+c}{(1+k)^{2}}+
\nonumber \\
&&\hspace*{-5mm}
\lim_{\tau\to\infty}\frac{1}{\tau}\sum_{u\leq\tau}
\sum_{t^\prime t^\pprime}
(\openone+\tilde{G})_{u+t~ t^\prime}^{-1}\tilde{C}_{t^\prime t^\pprime}(\openone+\tilde{G}^T)_{t^\pprime u}^{-1}
\label{eq:sigma2}
\end{eqnarray}
Clearly, the asymptotic (stationary)
value of the volatility $\sigma^2=\Xi(0)$
 cannot be
expressed in terms of persistent order parameters only.
It requires solving our
coupled saddle-point equations (\ref{eq:finalCandG}) for $C_{tt^\prime}$
and $G_{tt^\prime}$ for large times but finite
temporal separations $t-t^\prime$.
The persistent market correlations,
however, are found to be expressible in terms of persistent order
parameters:
\begin{equation}
\Xi(\infty)=
\frac{1+c}{2(1+k)^{2}}
\label{eq:asymp_correlations}
\end{equation}
In order to find the volatility we separate
the correlations at stationarity in a `frozen' and a `fickle' contribution:
\begin{eqnarray*}
C(t-t^\prime)&=&
\phi \bra \sgn[\tilde{q}(t)\tilde{q}(t^\prime)]\ket_{\rm fr}+
(1-\phi)\bra \sgn[\tilde{q}(t)\tilde{q}(t^\prime)]\ket_{\rm fi}
\\
&=&
\phi +
(1-\phi)\bra \sgn[\tilde{q}(t)]\sgn[\tilde{q}(t^\prime)]\ket_{\rm fi}
\end{eqnarray*}
and hence
\[
\tilde{C}(t-t^\prime)=
\phi-c +
(1-\phi)\bra \sgn[\tilde{q}(t)]\sgn[\tilde{q}(t^\prime)]\ket_{\rm fi}
\]
Insertion into (\ref{eq:sigma2}) and putting $t=0$ then gives
\begin{eqnarray}
2\sigma^2&=& \frac{1+\phi}{(1+k)^{2}}+
(1-\phi)\lim_{\tau\to\infty}\frac{1}{\tau}\sum_{t\leq\tau}
\sum_{t^\prime t^\pprime}
\nonumber \\
&&\times~
(\openone+\tilde{G})_{t t^\prime}^{-1}
\bra \sgn[\tilde{q}(t^\prime)]\sgn[\tilde{q}(t^\pprime)]\ket_{\rm fi}(\openone+\tilde{G}^T)_{t^\pprime t}^{-1}
\nonumber \\
&=& \frac{1+\phi}{(1+k)^{2}}+(1-\phi)
\nonumber \\[-1mm]
&&\times~
\lim_{\tau\to\infty}\frac{1}{\tau}\sum_{t\leq\tau}
\bra \left\{\sum_{t^\prime\leq t}(\openone+\tilde{G})_{t
t^\prime}^{-1}\sgn[\tilde{q}(t^\prime)]\right\}^2\!\ket_{\rm fi}
\label{eq:sigma3}
\end{eqnarray}
We note
that the sum
$\sum_{t^\prime<t}(\openone+\tilde{G})_{t
t^\prime}^{-1}\sgn[\tilde{q}(t^\prime)]$, is
the retarded self-interaction term in equation
(\ref{eq:singleagent}). Such a term is a familiar ingredient
of disordered systems with `glassy' dynamics (see e.g. \cite{FiscHert91}),
and generally acts as the mechanism
which drives the system to a `frozen' state.
Hence, self-consistency of the distinction between frozen
\begin{figure}[t]
\centerline{%
  \epsfxsize=80mm\epsfbox{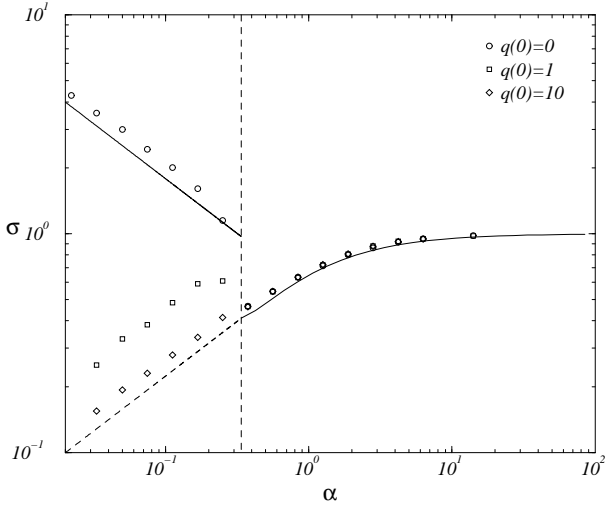}
}
\vspace*{3mm}
\caption{The volatility $\sigma$ as a function of the relative number $\alpha=p/N$ of
possible values for the external information.
The markers are obtained from individual simulation runs performed with a
system of $N=4000$ agents and various
homogeneous initial conditions, where $q_i(0)=q(0)$, and
in excess of 1000 iteration steps. The solid
curve for $\alpha>\alpha_c$ is the approximate expression
(\ref{eq:sigma4}). Below $\alpha_c$ the approximate asymptotic solutions
of Eqns. (\ref{eq:simplesigma}) (solid) and (\ref{eq:simplesigmafrozen}) (dashed) are drawn.}
\label{fig:volatility}
\end{figure}
\noindent
and
fickle traders dictates that the retarded self-interaction
term can be large for frozen traders, but must be
small (if not absent)  for fickle ones.
Our approximation now consists in consequently disregarding
the retarded self-interaction for the fickle traders:
\[
|\eta|<\frac{\sqrt{\alpha}}{1+k}:~~~~~~
\sum_{t^\prime<t}(\openone+\tilde{G})_{t
t^\prime}^{-1}\sgn[\tilde{q}(t^\prime)]\approx 0
\]
Thus we retain for fickle traders
only the instantaneous $t^\prime=t$ term in
$\sum_{t^\prime\leq t}(\openone+\tilde{G})_{t
t^\prime}^{-1}\sgn[\tilde{q}(t^\prime)]$, and find the (exact)
expression (\ref{eq:sigma3}) being replaced by the
approximation
\begin{equation}
\sigma^2= \frac{1+\phi}{2(1+k)^{2}}+\frac{1}{2}(1-\phi)
\label{eq:sigma4}
\end{equation}
This turns out to be a surprisingly accurate
approximation of the volatility for  $\alpha>\alpha_c$,
as can be observed in Fig. \ref{fig:volatility}.

Only in the limit $\alpha\to\infty$ can we expect to be able to
go beyond (\ref{eq:asymp_correlations}) and (\ref{eq:sigma4}), and work out
expressions (\ref{eq:sigma2}) and (\ref{eq:sigma3}) exactly. This requires
calculating the response function $\tilde{G}(\tau)$ for small
$\tau$, which we will set out to do next.
Since we assume absent anomalous response we may choose
trivial initial conditions. We also choose the perturbation
fields $\theta(t)$ to be non-zero only for a given time $t-\tau$, where $\tau>0$.
From (\ref{eq:singleagent}) we now derive
\begin{eqnarray}
\sgn[ q(t)] &=&\sgn\left[\frac{\theta(t-\tau)}{t\sqrt{\alpha}}
+ \frac{1}{t}\sum_{t^\prime\leq t} \eta(t^\prime)\right.
\nonumber \\
&&\left.- \frac{\sqrt{\alpha}}{t}\sum_{t^\prime t^\pprime\leq t}
   (\openone+ G)^{-1}_{t^\prime t^\pprime} {\rm sgn}[q(t^\pprime)]
\right]
\label{eq:largealpha_tool}
\end{eqnarray}
Hence, for vanishingly small perturbations $\theta(t-\tau)$, and
upon taking
the $t\to\infty$ limit:
\begin{eqnarray*}
\tilde{G}(\tau) &=&
 - \frac{2\sqrt{\alpha}}{1+k}
 \lim_{t\to\infty}\frac{1}{t}
\sum_{t^\prime \leq t}
\bra\delta\left[\eta- \frac{s\sqrt{\alpha}}{1+k}\right]
\left[
   \frac{\partial{\rm
   sgn}[q(t^\prime)]}{\partial \theta(t^\prime-\tau)}
  \right]\ket
\\
&&
+2\bra\delta\left[\eta- \frac{s\sqrt{\alpha}}{1+k}\right]
\left[
\lim_{t\to\infty}\frac{1}{t}\sum_{t^\prime\leq t}\frac{\partial\eta(t^\prime)}{\partial\theta(t-\tau)}
  \right]\ket
\end{eqnarray*}
We observe that $\eta=s\sqrt{\alpha}/(1+k)$ is precisely the
condition for a trader to be fickle, in the language of the
effective single agent. Secondly, from
causality it follows that
$\lim_{t\to\infty}t^{-1}\sum_{t^\prime\leq
t}\partial\eta(t^\prime)/\partial\theta(t-\tau)
=\lim_{t\to\infty}t^{-1}\sum_{t^\prime=t-\tau+1}^t
\partial\eta(t^\prime)/\partial\theta(t-\tau)=0$.
Hence our result can in a stationary state be written as
\begin{equation}
\tilde{G}(\tau) =
 - \frac{2\sqrt{\alpha}(1-\phi)}{1+k}
 \lim_{t\to\infty}
\bra \frac{\partial{\rm sgn}[q(t)]}{\partial \theta(t-\tau)}
\ket_{\rm fi}
\label{eq:response_explicit}
\end{equation}
For $\alpha\to \infty$ our stationary order parameter
equations give $(1-\phi)/(1+k)\to 1$.
 Furthermore, for $\alpha\to\infty$ all traders will become fickle, so
 $\bra \partial{\rm sgn}[q(t)]/\partial \theta(t-\tau)
\ket_{\rm fi}\to \tilde{G}(\tau)$. This leaves for $\alpha\to\infty$ only the trivial solution
for equation (\ref{eq:response_explicit}):
$\lim_{\alpha\to\infty}\tilde{G}(\tau)=0$ for all $\tau$.
Insertion into our exact expression (\ref{eq:sigma2}) for the
stationary volatility matrix
gives
\[
\lim_{\alpha\to\infty}\Xi(t)
=\frac{1}{2}
+\frac{1}{2}\lim_{\alpha\to\infty}\tilde{C}(t)
\]
and hence
\begin{equation}
\lim_{\alpha\to\infty}\lim_{t\to\infty}\sigma=1
\end{equation}
This is the random trading limit.

\section{The Stationary State For $\alpha<\alpha_c$}

\begin{figure}[t]
\centerline{%
  \epsfxsize=80mm\epsfbox{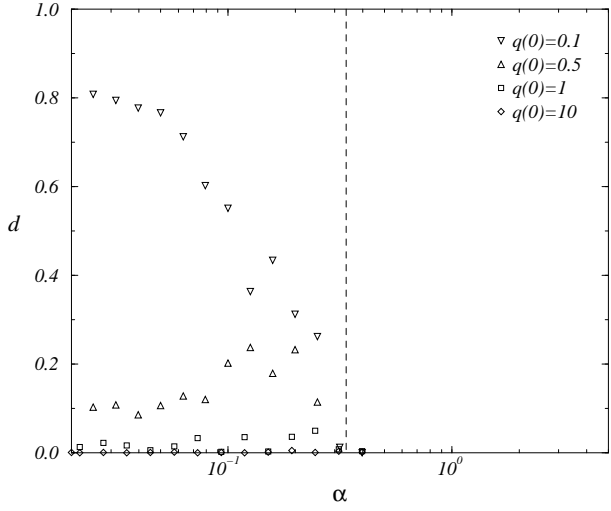}
}
\vspace*{3mm}
\caption{\label{fig:d}
The oscillatory component $d$ of the covariance (see equation (\ref{eq:ansatz_C})). The
markers represent the results of individual simulations, performed with $N=4000$ agents and various
homogeneous initial conditions, where $q_i(0)=q(0)$, and after in excess of 1000 iteration steps.}
\end{figure}

When the amount of external information available for agents to base their
actions upon (i.e. the value of $\alpha$), becomes small, the behaviour of the
market is found to become strongly dependent on initial conditions.
Numerical simulations show that below $\alpha_c$ the sequence $\sum_{t^\prime} G_{tt^\prime}$ is
unbounded, and that within the limits of experimental accuracy:
\begin{eqnarray}
 && \lim_{t\rightarrow\infty} \sum_{t^\prime}(\openone+G)^{-1}_{tt^\prime} = 0
\label{eq:ansatz_G}\\
&& C_{t+\tau, t}=c + d(-1)^{\tau}
  ~~~{\rm for}~~~\tau\neq 0
\label{eq:ansatz_C}
\end{eqnarray}
(with $C_{tt}=1$, by definition).
 Figure \ref{fig:d} shows the
asymptotic values of $d$ as measured during numerical simulations,
for different values of $\alpha$ and $q(0)$.
One clearly observes the dependence on initial
conditions.

We will now use (\ref{eq:ansatz_G},\ref{eq:ansatz_C}) as
{\em ans\"{a}tze}, i.e. we will construct special self-consistent
stationary state solutions of the
fundamental order parameter equations (\ref{eq:finalCandG}) which
obey (\ref{eq:ansatz_G},\ref{eq:ansatz_C}), as well as the
stationary state conditions $C_{t t^\prime}=C(t-t^\prime)$ and
$G_{tt^\prime}=G(t-t^\prime)$.
First we analyse the statistical properties of the Gaussian noise $\eta(t)$
in the single agent
equation (\ref{eq:singleagent}). From
(\ref{eq:ansatz_G},\ref{eq:ansatz_C}) it  follows that the noise
covariance matrix  (\ref{eq:noise_covariance}) obeys
\begin{eqnarray}
\lim_{t\to\infty}&&\bra \eta(t+\tau)\eta(t)\ket=
 (-1)^\tau d\gamma^2+
 \nonumber\\
 &&(1-c-d)\sum_{t}
 (\openone+G)^{-1}(t+\tau)(\openone+G)^{-1}(t)
 \label{eq:noisecov_smalla}
\end{eqnarray}
in which
\begin{equation}
\gamma=\sum_{t}(\openone+G)^{-1}(t)(-1)^{t}
\label{eq:gamma}
\end{equation}
From (\ref{eq:noisecov_smalla}) one can derive, in turn, that the
noise variables must asymptotically take the form:
\begin{equation}
t\to\infty:~~~
\eta(t)=(-1)^t\gamma z\sqrt{d}+\xi(t)\sqrt{1-c-d}
\label{eq:noise_smalla}
\end{equation}
where $z$ and $\{\xi(t)\}$ are zero-average Gaussian variables,
with $\bra z^2\ket=1$, $\bra z\xi(t)\ket=0$, and
\[
\lim_{t\to\infty} \bra \xi(t+\tau)\xi(t)\ket=
\sum_{t} (\openone+G)^{-1}(t+\tau)(\openone+G)^{-1}(t)
\]
Due to (\ref{eq:ansatz_G}) we know that $\lim_{\tau\to\infty}\lim_{t\to \infty}\bra
\xi(t+\tau)\xi(t)\ket=0$, i.e. in the stationary state the $\xi(t)$ decorrelate for large
temporal separations.
For sufficiently large $t$, and without external perturbations, equation (\ref{eq:singleagent})
now acquires the form
\begin{eqnarray}
  q(t+\!1) &=& q(t)+\gamma z\sqrt{\alpha d}(-1)^t
  +\xi(t)\sqrt{\alpha(1\!-c\!-d)}
\nonumber  \\
 && -\alpha \sum_{t^\prime\leq t} (\openone+ G)^{-1}_{t t^\prime} {\rm sgn}[q(t^\prime)]
  \label{eq:singleagent_largea}
\end{eqnarray}
Frozen agents are those for which $\sgn[q(t)]$ is independent
of time; due to (\ref{eq:ansatz_G}) these will not experience the
last term in (\ref{eq:singleagent_largea}).
However, due to the properties of the noise in the $\alpha<\alpha_c$
regime (and in contrast to the situation with $\alpha>\alpha_c$), even
frozen agents will now have
$\lim_{t\to\infty}q(t)/t= 0$.
Insertion into Eqn.
(\ref{eq:singleagent_largea}) shows that frozen solutions of the following form
 exist:
\begin{equation}
q(t)=q-\frac{1}{2}\gamma z\sqrt{\alpha d}(-1)^t
\label{eq:solution_frozen}
\end{equation}
provided $\sgn[q(t)]=\sgn[q]$ for all $t$, so $q$ and $d$ must obey
\begin{equation}
d=1-c,~~~~~~|q|>|\frac{1}{2}\gamma z\sqrt{\alpha d}|
\label{eq:condition_frozen}
\end{equation}
Oscillating agents, on the other hand, are those for which $\sgn[q(t)]=\hat{\sigma} (-1)^t$, with $\hat{\sigma}=\pm 1$.
Insertion into Eqn.
(\ref{eq:singleagent_largea}) shows that oscillating solutions of the following form
exist:
\begin{eqnarray}
q(t)=q+\frac{1}{2}\gamma\hat{\sigma}[\alpha- z\hat{\sigma}\sqrt{\alpha d}]
(-1)^t
\end{eqnarray}
provided $\sgn[q(t+1)]=-\sgn[q(t)]$ for all $t$, so $q$ and $d$ must obey
\begin{equation}
d=1\!-\!c,~~~~\gamma[\alpha\!- \!z\hat{\sigma}\sqrt{\alpha d}]>0,~~~~
|q|<\frac{1}{2}\gamma[\alpha\!-\! z\hat{\sigma}\sqrt{\alpha d}]
\label{eq:condition_oscill}
\end{equation}
Note that, if rigorously frozen and/or rigorously oscillating agents were to be
asymptotic solutions of
(\ref{eq:singleagent_largea}), then the correlations would have come
out as $C(\tau)=\phi+(1-\phi)(-1)^\tau$ (with $\phi$, as before, denoting the
fraction of frozen agents), and we would have found $c+d=1$.
Figures \ref{fig:c} and \ref{fig:d}, however, show that this
simple relation holds only near $\alpha=0$. Away from $\alpha=0$ there will therefore
 be solutions describing fickle agents which change strategy at
intervals intermediate between one (oscillating) and infinity
(frozen).
This can be understood on the basis of
(\ref{eq:singleagent_largea}), where due to the noise term $\xi(t)$ (with
a finite temporal correlation length) there will for $c+d<1$ always be a non-zero probability
of nearly frozen agents changing strategy occasionally, and of nearly
oscillating agents not changing strategy occasionally.

\section{The Limit $\alpha\to 0$}

Let us finally investigate the situation near $\alpha=0$ more
closely, where we may use the experimental observation
that $c+d\approx 1$, which implies that all agents will be either
frozen or oscillating. We put $c=\phi$ (the fraction of frozen agents) and
$d=1-\phi$, and choose homogeneous initial conditions with
$q(0)>0$. We now find
$\eta(t)=(-1)^t\gamma z\sqrt{(1-\phi)}$
and
our two solution types given by:
\begin{eqnarray*}
{\rm frozen}:&~~~ q(t)=&q-\frac{1}{2}\gamma z\sqrt{\alpha (1-\phi)}(-1)^t
\\
{\rm oscillating}:&~~~
q(t)=&
q+\frac{1}{2}\gamma\hat{\sigma}[\alpha- z\hat{\sigma}\sqrt{\alpha (1-\phi)}](-1)^t
\end{eqnarray*}
provided the following respective conditions for existence are met:
\begin{eqnarray}
{\rm frozen}:&&~~~
|q|>|\frac{1}{2}\gamma z\sqrt{\alpha (1-\phi)}|
\label{eq:cond_froz_new}
\\
{\rm oscillating}:&&~~~
|q|<\frac{1}{2}\gamma[\alpha- z\hat{\sigma}\sqrt{\alpha (1-\phi)}]
\nonumber \\
&&~~~\gamma\sqrt{\alpha}>\gamma z\hat{\sigma}\sqrt{1-\phi}
\label{eq:cond_osci_new}
\end{eqnarray}
Near $\alpha=0$ we also know, due to $c+d=1$, that
\begin{eqnarray}
t\to\infty: &&~~~
\bra \eta(t+\tau)\eta(t)\ket=(-1)^\tau (1-\phi)\gamma^2
\label{eq:noisecov_smallanew}\\
t\to\infty: &&~~~
\eta(t)=(-1)^t\gamma z\sqrt{1-\phi}
\label{eq:noise_smallanew}
\end{eqnarray}
and that
$\lim_{t\to\infty}\sigma^2=\frac{1}{2}(1-\phi)\gamma^2$.
In order to eliminate the remaining parameters $\gamma$ and $\phi$
we note that time translation invariance guarantees  the validity
of the relation
$\sum_{t}(G^n)(t)(-1)^{t}=[\sum_{t}G(t)(-1)^{t}]^n$, and hence
\begin{equation}
\gamma=(1+\Gamma)^{-1}~~~~~~~~\Gamma=\sum_{t}G(t)(-1)^{t}
\label{eq:gammas}
\end{equation}
The quantity $\Gamma$ can, in turn, be expressed in terms of $\gamma$ upon inserting
(\ref{eq:noisecov_smallanew},\ref{eq:noise_smallanew}) into
(\ref{eq:noise_relation}). We obtain
\[
\sqrt{\alpha}(1-\phi)\gamma(1-\gamma)(-1)^\tau=\lim_{t\to\infty}
\bra \sgn[q(t+\tau)]\eta(t)\ket_\star
\]
Working out the average in the right-hand side, by separating frozen from
fickle solutions, gives for large $t$:
\begin{eqnarray*}
\bra \sgn[q(t\!+\!\tau)]\eta(t)\ket_\star
&=&
\phi~\bra \sgn[q(t\!+\!\tau)]\eta(t)\ket_{\rm fr}\\
&&
+(1\!-\!\phi)~\bra \sgn[q(t\!+\!\tau)]\eta(t)\ket_{\rm fi}
\\
&=&
\gamma\sqrt{(1\!-\!\phi)}(-1)^{\tau}\left\{
\phi (-1)^t \bra \sgn[q] z\ket_{\rm fr}
\right. \\
&&\left.
~~~~~+
(1\!-\!\phi)~ \bra \hat{\sigma}  z\ket_{\rm fi}\right\}
\end{eqnarray*}
Since in a stationary state the correlation function
$\bra \sgn[q(t)]\eta(t^\prime)\ket_\star$ can only depend on
$t-t^\prime$, we must conclude that $\bra \sgn[q] z\ket_{\rm fr}=0$
and that either
\begin{equation}
\lim_{\alpha\to 0}\gamma(1\!-\!\phi)=0~~~~~~{\rm or}~~~~~~
\gamma=1\!-\!\sqrt{\frac{1\!-\!\phi}{\alpha}}~ \bra \hat{\sigma}  z\ket_{\rm fi}
\label{eq:gamma_found}
\end{equation}
(in leading order for $\alpha\to 0$).
Multiplication of both sides of the second equation in (\ref{eq:gamma_found}) by $\gamma\sqrt{\alpha}$ shows that
it automatically ensures the validity of
the second condition of (\ref{eq:cond_osci_new}).
The first equation of (\ref{eq:gamma_found}) will satisfy
the second condition of (\ref{eq:cond_osci_new}) as long as $\gamma>0$.

In order to proceed we need to calculate the persistent term $q$ in
the proposed solutions, which can be seen as representing their `effective initial conditions'.
It incorporates both the true initial
conditions and  the effects of the transients of the dynamics, which initially
will not be of the
simple form (\ref{eq:singleagent_largea}). Exact evaluation would
require solving our order parameter equations for arbitrary times,
which is not feasible. However, one can for now proceed on the
basis of the postulate that the properties of the long-term attractors (viz. the Gaussian
variable $z$) are uncorrelated with the value of $q$.
The conditions (\ref{eq:cond_froz_new},\ref{eq:cond_osci_new})
then simply state whether a value of $q$, generated independently of $z$ according to
some distribution $P(q)$,
is compatible with a given attractor.
Although we will not be able to generate {\em all} possible
stationary solutions of the process (\ref{eq:singleagent}), we
will show how two qualitatively different solutions, one with a diverging volatility for $\alpha\to 0$ and one with
a vanishing volatility for $\alpha\to 0$, can both be extracted from our
equations.

The first type of solution is obtained for $\lim_{\alpha\to
0}\phi=\phi_0<1$. Now one finds, in leading order in
$\alpha$, that $\hat{\sigma}=-\sgn[\gamma z]$ and that $\gamma=\bra |z|\ket_{\rm fi}\sqrt{(1-\phi_0)/\alpha}$.
The conditions (\ref{eq:cond_froz_new},\ref{eq:cond_osci_new})
reduce in leading order to the complementary pair
\begin{eqnarray}
{\rm frozen}:&&~~~
|q|>\frac{1}{2}\gamma |z|\sqrt{\alpha (1-\phi_0)}
\\
{\rm oscillating}:&&~~~
|q|<\frac{1}{2}\gamma |z|\sqrt{\alpha (1-\phi_0)}
\end{eqnarray}
This, in turn, allows us to calculate $\phi_0$ and $\bra |z|\ket_{\rm
fi}$:
\begin{eqnarray*}
\phi_0&=&\int\!dq~P(q)~\int\!\frac{dz}{\sqrt{2\pi}}e^{-\frac{1}{2}z^2}\theta\left[|q|-
\frac{1}{2}\gamma |z|\sqrt{\alpha (1-\phi)}\right]
\\
&=&\int\!dq~P(q)~\erf\left[\frac{\sqrt{2}|q|}{\gamma\sqrt{\alpha
(1-\phi)}}\right]
\end{eqnarray*}
\begin{eqnarray*}
\bra |z|\ket_{\rm fi}&=&
\int\!\frac{dq~P(q)}{1\!-\!\phi_0}\int\!\frac{dz~|z|}{\sqrt{2\pi}}e^{-\frac{1}{2}z^2}\theta\left[
\frac{1}{2}\gamma |z|\sqrt{\alpha (1\!-\!\phi_0)}\!-\!|q|\right]
\\
&=&
\frac{\sqrt{2}}{(1\!-\!\phi_0)\sqrt{\pi}}\int\!dq~P(q)~e^{-2q^2/\gamma^2\alpha (1-\phi_0)}
\end{eqnarray*}
We eliminate $\gamma$ in favour of $\sigma=\frac{1}{2}\sqrt{2}\gamma \sqrt{1-\phi_0}$
and end up with the following simple closed equation for $\sigma$:
\begin{equation}
\sigma
=
\int\!dq~P(q)~\frac{e^{-q^2/\sigma^2\alpha}}{\sqrt{\alpha\pi}}
\label{eq:almostfinalsigma}
\end{equation}
The associated value for $\phi_0$ then follows from:
\begin{equation}
\phi_0
=\int\!dq~P(q)~
\erf\left[\frac{|q|}{\sigma\sqrt{\alpha}}\right]
\label{eq:almostfinalphi}
\end{equation}
Finally we can use our observations
regarding the first few time-steps (section VI) of the process in order to obtain an estimate for $P(q)$.
These showed for small $\alpha$ that
initially
(i) for small $|q(0)|=\order(\sqrt{\alpha})$ the system is driven towards
the oscillating state, (ii) for large $|q(0)|=\order(\alpha^0)$ the
system tends to freeze, (iii) the transient processes are
dominated by the (Gaussian) noise term in (\ref{eq:singleagent}),
and
\begin{figure}[t]
\centerline{%
  \epsfxsize=85mm\epsfbox{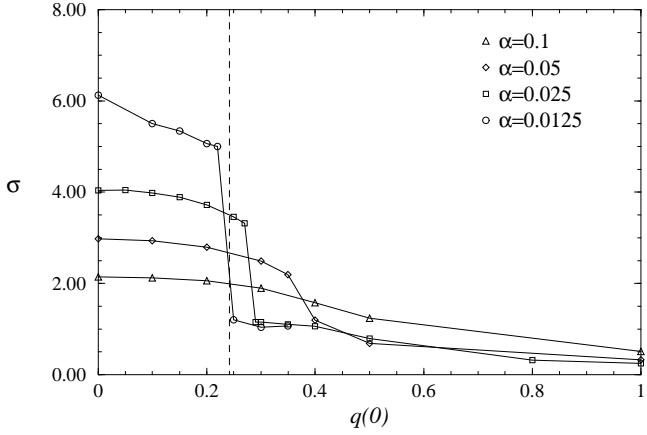}
}
\vspace*{3mm}
\caption{\label{fig:q_crit}
Experimental evidence in support of the existence of a critical value
for the initial strategy valuation $q(0)$ below
which a high-volatility solution exists.
The connected markers represent the results of measuring the volatility
in individual simulations, performed with $N=4000$ agents and initial conditions where $q_i(0)=q(0)$,
and after in excess of 1000 iteration steps.
CPU and memory limitations prevent us from doing reliable and
conclusive
experiments for $\alpha<0.0125$; the available data, however, are
clearly not in conflict with
our theoretical prediction
$q_c\approx 0.242$ (vertical dashed line), which follows from equation
(\ref{eq:transcendental}).}
\end{figure}
\noindent
(iv) the noise term is automatically  being `amplified' (either
via a diverging response function, or via accumulation over time)
to an effective $\order(\alpha^0)$
contribution.
Note that (i) and (ii) confirm that $q$ can indeed be
seen as the sum of $q(0)$ and the net effect of the transient
processes, and that (iii) and (iv) subsequently suggest
representing the transient processes by adding a single effective Gaussian
variable.
Hence for small $\alpha$ it would appear sensible to write
$P(q)=(\Lambda
\sqrt{2\pi})^{-1}e^{-\frac{1}{2}[q-q(0)]^2/\Lambda^2}$,
which converts (\ref{eq:almostfinalsigma},\ref{eq:almostfinalphi})
into
\[
\sigma^2\alpha+2\Lambda^2
=\frac{1}{\pi}
e^{-\frac{2q^2(0)}{\sigma^2\alpha+2\Lambda^2}}
\]
We conclude that $\sigma$ can be written in terms of the solution $y$ of a
transcendental equation
\begin{equation}
\sigma=\frac{1}{\sqrt{\alpha}}\left[\frac{2q^2(0)}{y}-2\Lambda^2\right]^{\frac{1}{2}}
~~~~~~~~
2q^2(0)=\frac{y}{\pi} e^{-y}
\label{eq:transcendental}
\end{equation}
For $|q(0)|\to 0$ we find that $\sigma=(\alpha\pi)^{-\frac{1}{2}}\sqrt{1-2\pi\Lambda^2}$,
hence we must obviously require $\Lambda^2<1/2\pi$.
The associated value for $\phi_0$ then follows from:
\begin{equation}
\phi_0
=\int\!Dx~
\erf\left[\frac{|q(0)+\Lambda x|}{\sigma\sqrt{\alpha}}\right]
\label{eq:ficklephi}
\end{equation}
Since we cannot calculate or estimate the width $\Lambda$
of the effective Gaussian noise term without solving our order parameter equations for short times
($\Lambda$ could even depend on $q(0)$),
it is quite satisfactory that
several interesting properties of the solution are found to be independent
of $\Lambda$. For instance, one always finds a diverging volatility of the form $\sigma=\order(\alpha^{-\frac{1}{2}})$,
and there is a critical value $q_c=(2\pi e)^{-\frac{1}{2}}\approx 0.242$ such that for $|q(0)|>q_c$
the solution no longer exists.
This solution is clearly the type of volatile state which has been reported regularly (see e.g.
\cite{SaviManuRiol99,ManuLiRiolSavi98})
upon observing numerical simulations. We have now found, however, that
whether or not it will
appear depends critically on the choice made for the initial conditions.
Numerical simulations indeed appear to support the existence and predicted magnitude of a
critical value $q_c\approx 0.242$  (see figure \ref{fig:q_crit}); fully conclusive
experiments, however (with even smaller values of $\alpha$),
would require impractical amounts of CPU and/or memory
in order to meet the requirements $p\to\infty$ and $N\to \infty$ for
increasingly small values of $\alpha$, and are presently ruled out.
In the limit $q(0)\to 0$ one can easily carry out the integrals in
(\ref{eq:ficklephi}), giving
$\Lambda=(2\pi)^{-\frac{1}{2}}\sin[\frac{1}{2}\pi\phi_0]$. Elimination of $\Lambda$
via insertion into $\sigma=(\alpha\pi)^{-\frac{1}{2}}\sqrt{1-2\pi\Lambda^2}$
then leads to the simple relation
\begin{equation}
\alpha,q(0)\to 0:~~~~~~~~~~
\sigma=\frac{\cos[\frac{1}{2}\pi
\phi_0]}{\sqrt{\alpha\pi}}+\order(\alpha^0)
\label{eq:simplesigma}
\end{equation}
This is the high-volatility solution shown in the $\alpha<\alpha_c$ regime of figure
\ref{fig:volatility}, with $\phi_0$ as measured in simulations
(see e.g. figure \ref{fig:frozen}). The power of $\alpha$ in (\ref{eq:simplesigma}) is
observed to be correct. The observed difference between theory and experiment
with regard to the prefactor can be understood as a
reflection of our approximation $c+d\approx 1$; this amounts to
disregarding deviations from the idealized `purely frozen' or
`purely oscillating' behaviour, which can indeed be expected to
give an approximate theory which (even for small $\alpha$) slightly under-estimates the
volatility.

We note that the condition $\lim_{\alpha\to 0}\phi<1$ for the above reasoning to apply
 can in fact be
weakened to $\lim_{\alpha\to 0}\alpha/(1-\phi)=0$.
The above solution ceases to hold, however, at the point where the fraction $\phi$ of frozen agents
scales as  $\phi=1-\kappa\alpha+\order(\alpha^2)$, in which case we
have to turn to the first option in
(\ref{eq:gamma_found}), rather than the second.
This is consistent with our previous observation that small values of
$|q(0)|$ lead to a relatively small fraction of frozen agents (and a large volatility),
whereas for large $|q(0)|$ such a solution will break down
in favour of states with a larger fraction of frozen agents.
Since we can now no longer use the second equation in
(\ref{eq:gamma_found}) to determine $\gamma$, and hence find
the volatility $\sigma=\frac{1}{2}\sqrt{2}\gamma\sqrt{1-\phi}$,
we have to return to (\ref{eq:gammas}). A fully frozen state,
which for $\alpha\to 0$ will indeed be described by this second type of solution
(since $\lim_{\alpha\to 0}\phi=1$), must necessarily have
$G(t>0)=g$. This is consistent with our {\em ans\"{a}tze}, since
it gives
\[
t>0:~~~~~~
(\openone+G)^{-1}(t)=
-g(1-g)^{t-1}
\]
which implies $\sum_{t\geq 0}(\openone+G)^{-1}(t)=0$, provided
$0<g<2$. We can now calculate $\gamma$ from (\ref{eq:gammas}) and
find $\lim_{\alpha\to 0}\gamma=2/(2-g)$.
Thus we obtain, provided $2-g=\order(\alpha^0)$:
\begin{figure}[t]
\centerline{%
  \epsfxsize=85mm\epsfbox{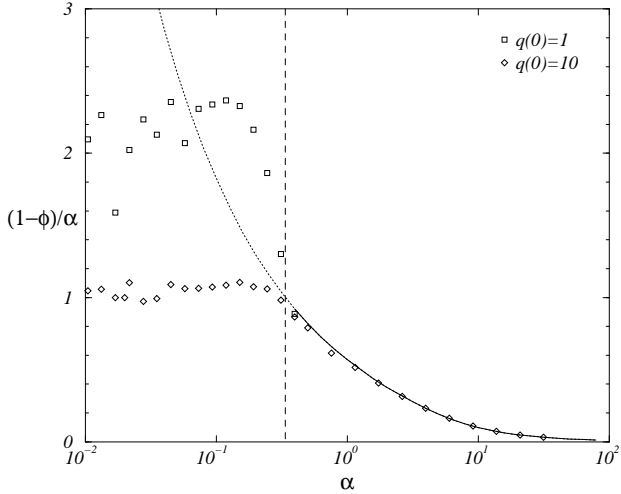}
}
\vspace*{3mm}
\caption{\label{fig:smallalphalimit}
Experimental evidence for the existence of the limit $\kappa=\lim_{\alpha\to 0}(1-\phi)/\alpha$
for the low-volatility solution. The
markers are obtained from individual simulation runs performed with a system of $N=4000$
agents and initial valuations of the form $q_i(0)=q(0)>q_c$ (to evoke the low-volatility state),
and in excess of 1000 iteration steps.
The solid curve to the right of the
critical point is the theoretical prediction, obtained from
the exact equations (\ref{eq:c}) and $\phi=1-\erf[\sqrt{\alpha/2(1+c)}]$
describing the $\alpha>\alpha_c$ regime.
The dotted curve to the left is its continuation into the $\alpha<\alpha_c$ regime
(where it should indeed no longer be correct).}
\end{figure}
\noindent
\[
\sigma=\frac{\sqrt{2\kappa}}{2\!-\!g}\sqrt{\alpha}+\order(\alpha)
~~~~~~~~
\kappa=\lim_{\alpha\to 0}\frac{1\!-\!\phi}{\alpha}
\]
We also note that the scaling property $\phi=1-\order(\alpha)$
implies that $P(0)=\lim_{q\to 0}P(q)=\order(\sqrt{\alpha})$, since
all $q$ values of order
$q=\order(\sqrt{\alpha})$ will contribute to the fraction $1-\phi$ of fickle
agents, giving $1-\phi=\order(P(0)\sqrt{\alpha})$.
We can now calculate $\lim_{\alpha\to 0}g$ upon explicitly inspecting
the effect of a perturbation of a frozen state. In view of $G(t>0)=g$
we may restrict ourselves to considering the effect on $\sgn[q(t+1)]$
of a perturbation at time $t$, giving in leading order for $\alpha\to
0$:
\begin{eqnarray*}
\lim_{\alpha\to 0}g&=&\lim_{\alpha\to 0}\lim_{\theta\to 0}\bra \frac{\partial}{\partial\theta}~\sgn\left[
q+\frac{1}{2}\alpha \gamma z\sqrt{\kappa}(-1)^t
+\theta
 \right]\ket
\\
&=&
2\lim_{\alpha\to 0}\bra \delta\left[
q+\frac{1}{2}\alpha \gamma z\sqrt{\kappa}(-1)^t
 \right]\ket
\\
&=&2\lim_{\alpha\to 0}P(0)=0
\end{eqnarray*}
Hence, since the frozen state has $q=\order(\alpha^0)$, we find $\lim_{\alpha\to 0}\gamma=1$ and
\begin{equation}
\alpha\to 0:~~~~~~~~
\sigma=\frac{1}{2}\sqrt{2\kappa \alpha}+\order(\alpha)
\label{eq:simplesigmafrozen}
\end{equation}
Explicit calculation of the prefactor in
(\ref{eq:simplesigmafrozen}) would require taking our calculations beyond the
leading order in $\alpha$, in order to determine to find $\kappa$.
Equation (\ref{eq:simplesigmafrozen})
is the low-volatility solution shown in the $\alpha<\alpha_c$ regime of figure
\ref{fig:volatility}, with $\kappa$ as measured in simulations
(see e.g. figure \ref{fig:smallalphalimit}). Again the power of $\alpha$ in (\ref{eq:simplesigmafrozen}) is
observed to be correct. The remaining difference between theory and experiment
with regard to the prefactor can again be understood as a
reflection of our approximation $c+d\approx 1$, which induces a structural under-estimation of the
volatility.

\section{Discussion}

In this paper we have solved a `batch' version of the
 minority game with random external information, using
generating functional analysis (or dynamic mean field theory)
\'{a} la De Dominicis, which allows one to carry out the disorder averages in a dynamical context.
Since the dynamics of the game is not described by a detailed balance type stochastic process, equilibrium
statistical mechanical tools can not be applied directly. Phase
transitions (if present) must be of a dynamical nature.
 The disorder in the minority game consists of the microscopic realisation of the
 repertoire of randomly drawn trading strategies of the $N$ agents.  Upon taking the limit $N\to\infty$
 (where $N$ denotes the number of agents playing the game) one ends up with an exact non-Markovian
stochastic  equation describing the dynamics of an effective single agent (\ref{eq:singleagent}),
whose statistical properties are identical to those of the
original system (averaged over all realisations of the disorder).
The key control parameter in this problem is the ratio $\alpha=p/N$
of the number of possible values of the external information over
the number of agents.

We find a phase transition at $\alpha_c=0.33740$, signaled by the
onset of anomalous response, in agreement with the value reported
recently in \cite{ChalMarsZecc00}. The method used in
\cite{ChalMarsZecc00} depends on the fact that for their stochastic
version of the minority game a Lyapunov function exists. Our approach
does not have this constraint and can be easily applied to those
variations of the game where a Lyapunov function is not available,
thus opening up a wider range of models for analysis (see
e.g. \cite{Chalweb}).  Above $\alpha_c$ (where anomalous response is
absent) we can solve the stationary state of the system exactly,
giving exact expressions for quantities such as the fraction of frozen
agents (which is zero for $\alpha\to\infty$ but increases with
decreasing $\alpha$), the persistent two-time correlations, and the
persistent correlations in the total bid. The volatility (which is
itself not an order parameter of the system) can be calculated to a
very good approximation. Above $\alpha_c$, our method and that of
\cite{DemaMars00,ChalMarsZecc00} are likely to describe the same
behaviour \cite{privMars}.  Below $\alpha_c$, i.e. in the region of
complex dynamics (inaccessible by the replica approach
\cite{MarsChal01}), our present method still applies.  In this region
we demonstrate the existence of multiple stationary states, and derive
expressions for the relevant observables in leading order in $\alpha$
as $\alpha\to 0$.  We show, more specifically, that the occurrence and
practical observability of a diverging volatility for $\alpha\to 0$
(as reported in e.g.  \cite{SaviManuRiol99,ManuLiRiolSavi98}) is
crucially dependent on the overall degree of {\em a priori} preference
for specific strategies exhibited by the agents at $t=0$, which may
explain the different observations regarding the $\alpha\to 0$
behaviour which have been reported in literature
\cite{GarrMoroSher00}.  More specifically, our theory points at the
existence of a critical value for the initial strategy valuations,
above which the system will revert to a state with vanishing
volatility.  Our theoretical predictions find quite satisfactory
confirmation in numerical simulations.

The fact that we can analyse the stationary state of
Eqn. (\ref{eq:singleagent}), in spite of it describing a non-Markovian
stochastic process, suggests that the present method should also be
suitable to deal with models where the external information depends on
time, or on the previous behaviour of the agents, as in the original
model \cite{ChalZhan97}.

\acknowledgements We would like to thank Andrea Cavagna and David
Sherrington for introducing us to the Minority Games and Matteo
Marsili for a discussion of previous work done.  JAFH wishes to thank
King's College London Association for financial support.


\appendix

\section{Expressions for Average Bid and Volatility}
\label{sec:average_and_volatility}

First we calculate $\lim_{N\to\infty}\overline{\bra A_t\ket}$
using expression (\ref{eq:average}).
We note that we obtain $\bra A_t\ket$ simply by making the replacement
$e^{i\sum_{ti}\psi_i(t)q_i(t)}\to (\tau/\alpha N)\sum_\mu x_t^\mu$ in
the right-hand side of Eqn. (\ref{eq:Zbeforeaverage}).
The disorder average is carried out as before, but instead of Eqn. (\ref{eq:Zafteraverage})
we now obtain
\begin{eqnarray*}
\overline{\bra A_t\ket}&=&
 \tau \int\![DC D\hat{C}][DK D\hat{K}][ DL D\hat{L}]~
    e^{N\left[\Psi+\Phi+\Omega \right]+\order(N^0)}
\\
&&
\times ~e^{-\Phi/\alpha}\!\!
 \int\!\! Dw D\hat{w} Dx D\hat{x}~x_{t}^1~
  e^{i\sum_{s}[\hw_s w_s + \hx_s x_s +  w_s x_s]}
\nonumber\\
&&\times
\left.
e^{-\frac{1}{2}\sum_{s s^\prime} \left[
 w_s w_{s^\prime}
 +\hw_s L_{s s^\prime} \hw_{s^\prime}
 +2\hx_{s} K_{s s^\prime} \hw_{s^\prime}
 +\hx_s C_{s s^\prime}\hx_{s^\prime}
\right]}\right]
\end{eqnarray*}
where we have used permutation invariance with respect to $\mu$
(after the disorder average). The integral is dominated by the familiar saddle-point.
Since the $\order(N^0)$ term in the
exponent is identical to that in (\ref{eq:Zafteraverage}), we can
now simply use the identity $\overline{Z[{\bf 0}]}=1$ to show that
\[
~~
\lim_{N\to\infty}\overline{\bra A_t\ket}
\hspace*{\fill}\vspace*{-4mm}
\]
\begin{eqnarray}
~~~~&=&\tau e^{-\Phi/\alpha}
 \int Dw D\hat{w} Dx D\hat{x}~x_{t}~
  e^{i\sum_{s}[\hw_s w_s + \hx_s x_s +  w_s x_s]}
\nonumber\\
&&\times
\left.
e^{-\frac{1}{2}\sum_{s s^\prime} \left[
 w_s w_{s^\prime}
 +2i\hx_{s} G_{s s^\prime} \hw_{s^\prime}
 +\hx_s C_{s s^\prime}\hx_{s^\prime}
\right]}\right]
\nonumber\\ &=& 0
\label{eq:average_expression}
\end{eqnarray}
The last step follows immediately from the anti-symmetry
of the integrand under overall reflection.

To determine the disorder-averaged volatility matrix, which for $N\to\infty$
becomes identical to $\overline{\bra A_t A_{t^\prime}\ket}$ due to
(\ref{eq:average_expression}) and the self-averaging property,
we first work out
the dominant terms in (\ref{eq:volatility_matrix}). Using $\lim_{N\to\infty}(\alpha N)^{-1}\sum_\mu
\Omega_\mu^2=\frac{1}{2}$, we obtain the relatively simple
expression
\[
\lim_{N\to\infty}\bra A_t A_{t^\prime}\ket=\lim_{N\to\infty}\frac{1}{2\alpha
N} \sum_\mu \bra [x_t^\mu+\Omega^\mu/\tau][x_{t^\prime}^\mu+\Omega^\mu/\tau]\ket
\]
We calculate this average by making the replacement
$e^{i\sum_{ti}\psi_i(t)q_i(t)}\to
(2\alpha N)^{-1} \sum_\mu \bra [x_t^\mu+\Omega^\mu/\tau][x_{t^\prime}^\mu+\Omega^\mu/\tau]\ket$ in
the right-hand side of Eqn. (\ref{eq:Zbeforeaverage}).
Repeated integration by parts over the $w_t^\mu$ shows that we may equivalently put
$e^{i\sum_{ti}\psi_i(t)q_i(t)}\to (2\alpha N)^{-1}\sum_\mu
\hw_t^\mu \hw_{t^\prime}^\mu$.
Following the steps we also took in calculating $\overline{\bra
A\ket}$
now gives
\[
~~
\lim_{N\to\infty}\overline{\bra A_t A_{t^\prime}\ket}
\hspace*{\fill}\vspace*{-4mm}
\]
\begin{eqnarray}
~~~~&=& \frac{1}{2}e^{-\Phi/\alpha}\!\!
 \int\!\! Dw D\hat{w} Dx D\hat{x}~\hw_{t}\hw_{t^\prime}
  e^{i\sum_{s}[\hw_s w_s + \hx_s x_s +  w_s x_s]}
\nonumber\\
&&\times
\left.
e^{-\frac{1}{2}\sum_{s s^\prime} \left[
 w_s w_{s^\prime}
 +2i\hx_{s} G_{s s^\prime} \hw_{s^\prime}
 +\hx_s C_{s s^\prime}\hx_{s^\prime}
\right]}\right]
\nonumber\\
&=&
\frac{1}{2}
\frac{\int\!D\hw~\hw_t\hw_{t^\prime}~e^{-\frac{1}{2}\sum_{ss^\prime}\hw_s\left[
(\openone+G)^T D^{-1}(\openone+G)\right]_{ss^\prime}\hw_{s^\prime}}}
{\int\!D\hw~e^{-\frac{1}{2}\sum_{ss^\prime}\hw_s\left[
(\openone+G)^T D^{-1}(\openone+G)\right]_{ss^\prime}\hw_{s^\prime}}}
\nonumber\\
&=&\frac{1}{2}[(\openone+G)^{-1}D(\openone+G^T)^{-1}]_{t t^\prime}
\label{eq:volatility_expression}
\end{eqnarray}

\section{Consequences of Absence of Anomalous Response}

\begin{lemma}
Consider two bounded sequences of real numbers $A_t$ and $b_t$.
Because $b_t$ is bounded, there exists a number $b$ such
that $\lim_{\tau\rightarrow\infty}\frac{1}{\tau}\sum_{t\leq
\tau}b_t = b$. Define $a_\tau=\sum_{t\leq \tau}A_{t}$,
and assume that $\lim_{\tau\rightarrow \infty} a_\tau=a$. Then
\[
  \lim_{\tau\rightarrow \infty} \frac{1}{\tau}\sum_{t\leq \tau} \sum_{t^\prime\leq t}
  A_{t-t^\prime}b_t^\prime
  = a b
\]
\end{lemma}
\begin{proof}
Upon substituting of $t\to t+t^\prime$ we find
\[
  \frac{1}{\tau} \sum_{t\leq \tau} \sum_{t^\prime\leq t}
  A_{t-t^\prime}b_{t^\prime}
  = \frac{1}{\tau} \sum_{t^\prime\leq \tau} b_{t^\prime}
  \sum_{t\leq \tau-t^\prime}A_t
  =  \frac{1}{\tau} \sum_{t\leq \tau} a_{\tau-t} b_t
\]
The sequences $\{a\}$ and $\{b\}$ are bounded, so there exist
 numbers $C_a$ and $C_b$ such that $|a_t|<C_a$ and $|b_t|<C_b$ for all
$t\geq 0$. The sequence $\{a\}$ converges to $a$, so for any $\epsilon>0$
there exists an $K$ such that for all $t>K$:
$|a_t-a|<\epsilon/3C_b$. We now choose $M$ such that for all $\tau>M$,
$|\frac{1}{\tau}\sum_{t\leq \tau}b_t - b|<\epsilon/3|a|$ and
$K C_a C_b/ \tau<\epsilon/3$. Then we find for all $\tau>M$:
\begin{eqnarray*}
   \left|\frac{1}{\tau}\sum_{t\leq \tau} a_{\tau-t} b_t -
   a b\right|
  &=&
  \left|\frac{1}{\tau}\!\!\sum_{t=\tau-K}^\tau\!\! a_{\tau-t} b_t
  +\frac{1}{\tau}\!\!\sum_{t<\tau-K}\!\! a_{\tau-t} b_t - a b\right|
\end{eqnarray*}
  \vspace*{-5mm}
\begin{eqnarray*}
  &\leq&
  \left|\frac{1}{\tau}\!\!\sum_{t=\tau-K}^\tau\!\! a_{\tau-t} b_t\right|
  \\
  &&
  +\left|\frac{1}{\tau}\!\!\sum_{t<\tau-K}\!\! (a_{\tau-t}-a) b_t -
      a \left(b-\frac{1}{\tau}\!\!\sum_{t<\tau-K}\!\! b_t\right)\right|
  \\
~~  &\leq&
  \frac{K C_a C_b}{\tau}
  + \left|\frac{1}{\tau}\!\sum_{t<\tau-N}\! (a_{\tau-t}-a) b_t\right|
  + |a|\left|b-\frac{1}{\tau}\!\sum_{t<\tau-K}\! b_t\right|
  \\
  &\leq&
  \epsilon
\end{eqnarray*}
Hence the limit is as claimed.
\end{proof}
\begin{lemma}
Suppose $G_{st}=G(s-t)\in\Re$, where $G(t)=0$ for all $t<0$ and
with $\lim_{\tau\to\infty}\sum_{t\leq
\tau}G(t)=k$, and suppose
$\lim_{\tau\to\infty}\tau^{-1}\sum_{t\leq \tau}s(t)=s$. Then for
all $n\in\NN$:
\[
 \lim_{\tau\to\infty}\frac{1}{\tau}\sum_{t=1}^\tau
 \sum_{t^\prime}(G^n)_{tt^\prime}s(t^\prime)
  =
  k^n s.
\]
\end{lemma}
\begin{proof}
The proof proceeds by induction. For $n=0$, the statement is trivially true.
Suppose now that it is true for all $n\leq m$. Then
\[
 \lim_{\tau\to\infty}\frac{1}{\tau}\sum_{t=1}^\tau
 \sum_{t^\prime}(G^{m+1})_{tt^\prime}s(t^\prime)
  =~~~~~~~~~~~~~~~~~~~~~~
  \vspace*{-5mm}
\]
\[
~~~~~~~~~~~~~~~~
   \lim_{\tau\to\infty}\frac{1}{\tau}\sum_{t=1}^\tau
 \sum_{t^\pprime\leq t}G(t-t^\pprime)\sum_{t^\prime\leq t^\pprime}
 (G^{m})_{t^\pprime t^\prime}s(t^\prime)
\]
The sequence $b_t=\sum_{t^\prime\leq t}(G^m)_{t t^\prime}s(t^\prime)$
satisfies the conditions of the preceding
lemma, application of which gives
\[
 \lim_{\tau\to\infty}\frac{1}{\tau}\sum_{t=1}^\tau
 \sum_{t^\prime}(G^{m+1})_{tt^\prime}s(t^\prime)
 =k.k^{m}s=k^{m+1}s
\]
Hence the claim holds for $m+1$, and
by induction it is now proved for all $n$.
\end{proof}
\vspace*{\fill}

\end{document}